# Distinguishing Feedback Mechanisms in Clock Models

Alexander D. Golden[1], Joris Paijmans[2], and David K. Lubensky[1]

[1]*Department of Physics, University of Michigan, Ann Arbor MI 48109-1040, USA*

[2]*Max Planck Institute for the Physics of Complex Systems, Nöthnitzer Str. 38, D-01187 Dresden, Germany*

## Abstract

Biological oscillators are very diverse but can be classified based on dynamical motifs such as the types of feedback loops present. The *S. Elongatus* circadian clock is a remarkable phosphorylation-based oscillator that can be reconstituted *in vitro* with only 3 different purified proteins: the clock proteins KaiA, KaiB, and KaiC. Despite a growing body of knowledge about the biochemistry of the Kai proteins, basic questions about how their interactions lead to sustained oscillations remain unanswered. Here, we compare models of this system that make opposing assumptions about whether KaiA sequestration introduces a positive or a negative feedback loop. We find that the two different feedback mechanisms can be distinguished experimentally by the introduction of a protein that binds competitively with KaiA. Understanding the dynamical mechanism responsible for oscillations in the Kai system may shed light on the broader question of what clock architectures have been selected by evolution and why.

## Introduction

Circadian clocks are found in many organisms and are thought to provide advantages by synchronizing biological processes with the earth's day/night cycle. These time-keeping systems are biological oscillators capable of being entrained by periodic driving (like the daily alternation of light and dark) and of sustaining a robust period near 24 hours in the absence of external signals. Biological activities as diverse as metabolism [2] and gene regulation[3] have been shown to depend on circadian rhythms. Studies in various model systems have led to a number of conjectures about the consequences of different feedback architectures for clock performance. For example, some authors [4, 5] have suggested that the presence of positive feedback loops in biological oscillators could make oscillations simpler to achieve, while others have argued that additional negative feedback loops can provide robustness advantages [6] or that it is most important to maintain the correct balance between positive and negative feedbacks [7]. A major hurdle to fully assessing these proposals is that it is often difficult to determine the complete network structure of real biological clocks, even in the most tractable systems. Here, we address this problem for a particularly simple circadian oscillator, that belonging to the cyanobacterium *Synechococcus elongatus.* We identify experimentally accessible signatures of the feedback structures present in different mathematical models of this system and, in particular, show how one can determine whether the *in vitro* Kai clock derived from *S. elongatus* relies on a strong positive feedback loop. Our findings both clarify the clock architecture in an important model organism and shed light on generic design principles for sustaining reliable biological rhythms.

The *S. Elongatus* circadian clock is built around a core post-translational protein phosphorylation cycle that has the striking property that it can be reproduced *in vitro* with purified proteins [8, 9]. It consists of the 3 proteins KaiA, KaiB, and KaiC [10], which form complexes rhythmically in the presence of ATP. KaiC forms a homohexamer in the presence of ATP, undergoing cyclic phosphorylation and dephosphorylation on a threonine and then a serine residue on each KaiC monomer [11-15]. Over the course of an oscillation cycle, dimeric KaiA first binds to the C-terminal domain of unphosphorylated KaiC, promoting its phosphorylation. This phosphorylation induces conformational changes in the KaiC hexamer which promote KaiB binding to the KaiC N-terminal domain [13, 14, 16-18]. The KaiBC complex can then bind and sequester KaiA, preventing it from inducing autophosphorylation in other KaiC [19]. In the absence of free KaiA, hexamers dephosphorylate. The unphosphorylated hexamers unbind the KaiB, releasing the KaiA from sequestration. These KaiA are then free to promote phosphorylation of KaiC hexamers, thus restarting the cycle (Figure 1).

This cycle of protein interactions has been previously modeled in different ways [20-23]. Here we study two different proposed models for the Kai system that both focus on the importance of KaiA sequestration. We show that, even though they assume similar molecular interactions, they have very different mathematical properties that can be experimentally distinguished. We call these models the *allosteric* and *monomer* models due to the focus of the first model on the roles of different KaiC hexamer conformations and of the second on the multiple possible phosphorylation states of individual KaiC monomers.

These two models of the Kai system can be understood as representing examples of two qualitatively different mechanisms for generating oscillations in biological systems. The allosteric model, relying on negative feedback [24], is an example of a delay-based oscillator. In these oscillators a time delay and negative feedback drive the system past steady state, generating oscillations (Figure 2). Delays can come from many different sources, from protein synthesis times to long chains of intermediate reactions [25].

The monomer model [11] is an example of a "relaxation oscillator" [26] that relies on strong positive feedback (Figure 3a). The positive feedback causes the system to "overshoot" an unstable steady state, effectively switching between two slowly evolving states. This action can be understood in analogy to the delay oscillator: whereas the delay oscillator used a dependence on the previous state of the system to prevent the system from settling at a steady state, a relaxation oscillator achieves a similar effect with hysteresis, which keeps the system moving past the steady state. It also exhibits a strong separation of timescales: the system evolves along one of the slow states until it reaches a turning point and then, much more quickly, switches to the other slow state (Figure 3b).

The question of whether the *in vitro* Kai system is best described as a delay or a relaxation oscillator has yet to be resolved experimentally, as current results appear to provide contradictory evidence. The basic issue is whether KaiA bound to the KaiC C-terminal is only active during the phosphorylation phase of the oscillation or also effectively retards dephosphorylation. Even though the *in vitro* Kai system appears to be quite simple by the standards of biological clocks, it is still difficult to measure all of the relevant rate constants directly, and different approaches to estimating their values do not agree. In particular, experiments in which phosphomimetics are used to isolate certain reactions by fixing the phosphorylation state of one of the two residues that can accept phosphates [15, 27] seem to conflict with studies in which rates are instead inferred from fitting a kinetic model with multiple reactions to phosphorylation time courses of the native protein [11].

Here we show how the different possibilities for the type of feedback caused by KaiA sequestration can be distinguished experimentally without direct measurement of microscopic rate constants. We first introduce the models in detail and describe the distinct assumptions they make about the form of the feedback introduced by KaiA sequestration. We next show that the allosteric model and the monomer model exhibit opposite responses in both amplitude and period to changes in the efficiency of KaiA sequestration by the KaiB complex. These responses can be understood as consequences of the type of feedback each model exhibits. We then show that such changes in sequestration efficiency can be generated experimentally by a protein that competes with KaiA for binding on the KaiC N-terminal domain in the KaiBC complex. Recent research indicates that CikA is a strong candidate for this role [28]. Addition of CikA to the *in vitro* oscillator results in a decreased period [29], consistent with the results for the allosteric model. We finally show that the same qualitative behavior is seen in extensions of the basic allosteric and monomer models which maintain the same fundamental feedback structure [30, 31], including recent models that aim to provide a more detailed description of the biochemistry of the Kai proteins [32]. Varying efficiency of KaiA sequestration thus provides a robust way to directly probe whether the Kai oscillator is closer to a delay oscillator or to a relaxation oscillator.

# Models

In this section we describe the allosteric and monomer models and distinguish their feedback structures. In particular, we define the sequestration efficiency $m$ as the amount of KaiA sequestered per KaiC (per KaiC hexamer for the allosteric model and per KaiC monomer for the monomer model); this parameter will play a central role in our analysis of each of the following models. In the interest of showing that the principles valid for these comparatively simple models hold in a more realistic setting we then introduce a third model, the two-site allosteric model[32], that attempts to more faithfully capture the biochemical complexities of the full system.

## Allosteric model

The allosteric model (Figure 4a), introduced in [24], takes hexameric KaiC as its fundamental object. It combines the two phosphorylation sites on each monomer into one lumped site and assumes that each hexamer exists in one of two different allosteric states called *active* and *inactive* in analogy with the Monod-Wyman-Changeux model of conformational transitions. The transition rates between different conformations are assumed to depend on the number of phosphorylated monomers in a hexamer, with more phosphorylated hexamers preferring the inactive state and less phosphorylated hexamers preferring the active state. As the system evolves the population of active hexamers becomes sequentially more phosphorylated until the inactive conformation is preferred. Once in the inactive conformation the population then begins to dephosphorylate until it switches back to the active conformation. The large number of elementary steps in each of these processes produces the delay that is at the core of the oscillator.

In the allosteric model a KaiA monomer can bind to both active and inactive KaiC. KaiA binds to active KaiC and promotes autophosphorylation before unbinding. After a KaiC hexamer has changed conformation to the inactive state it can form a complex with two KaiB dimers which in turn binds $m$ KaiA dimers, thereby sequestering KaiA and preventing it from promoting phosphorylation. The inactive KaiC then dephosphorylates and begins to switch to the active state, at which point it begins to release the sequestered KaiA. When enough KaiA is free it induces the active KaiC to autophosphorylate until the inactive state is preferred again, completing the cycle. Thus the net effect of KaiA sequestration by inactive KaiC in the allosteric model is a negative feedback with a delay (Figure 2), preventing active KaiC from phosphorylating and retarding the progression of the cycle until dephosphorylation is complete. This model is described by the following chemical reactions [24], with mass action kinetics:

$$C_i \leftrightharpoons \tilde{C}_i \tag{1}$$

$$C_i + A \leftrightharpoons AC_i \rightarrow C_{i+1} + A \tag{2}$$

$$\tilde{C}_i + 2B \leftrightharpoons B_2\tilde{C}_i, \quad B_2\tilde{C}_i + mA \leftrightharpoons A_m B_2\tilde{C}_i \tag{3}$$

$$C_i \leftrightharpoons C_{i+1}, \quad \tilde{C}_i \leftrightharpoons \tilde{C}_{i+1} \tag{4}$$

$$B_2\tilde{C}_i \leftrightharpoons B_2\tilde{C}_{i+1}, \quad A_m B_2\tilde{C}_i \leftrightharpoons A_m B_2\tilde{C}_{i+1}. \tag{5}$$

Here $C_i$ represents a KaiC hexamer in the active state with $i$ phosphorylated monomers ($i$ ranging from 0 to 6) and $\tilde{C}_i$ represents a KaiC hexamer in the inactive state. $A$ and $B$ stand for KaiA and KaiB, respectively, and $m$ is the KaiA sequestration stoichiometry. Although $m$ would normally be an integer, velow we will sometimes take it to be a continuously varying real number, and extend the mathematical equations of deterministic mass action kinetics to this case, as described in the Supporting Material (SM). Since the Kai oscillation is most commonly understood as a phosphorylation oscillation, we will often consider the quantity $p(t)$, the phosphorylation fraction as a function of time. This is the proportion of KaiC monomers that are phosphorylated, and is defined by

$$p(t) = \frac{1}{C_T}\sum_{i=0}^{6} i\left(C_i + \tilde{C}_i + A_m B_2\tilde{C}_i + B_2\tilde{C}_i + A_m B_2\tilde{C}_{i+1}\right), \tag{6}$$

where $C_T$ represents the total concentration of KaiC hexamers. Here and throughout we use the same symbol for both a chemical species and its concentration and let context distinguish them. Unless otherwise stated, the parameters used for simulations of the allosteric model are those found in table S2 of [24].

## Monomer model

The monomer model (Figure 4b), proposed in [11], takes the individual KaiC monomer as its basic unit. It relies on ordered phosphorylation on the two residues (serine and threonine) that are known to have a key contribution to the circadian oscillation. In this model, if all KaiC monomers begin in the fully unphosphorylated state $U$ , first the threonine residue is phosphorylated, then the serine is phosphorylated yielding a doubly phosphorylated $D$ monomer, then the threonine is dephosphorylated and finally the serine is dephosphorylated. This leads to the following cycle describing a full oscillation: $U \to T \to D \to S \to U$ .

In the monomer model the presence of free KaiA directly alters the rates of each phosphorylation reaction as shown below. High free KaiA concentration promotes phosphorylation and low free KaiA concentration promotes dephosphorylation. KaiA binding is not explicit; instead, it is presumed that any free KaiA binds to $S$ form KaiC until one or the other of the species is entirely depleted. Thus, as the amount of $S$ KaiC increases it sequesters more and more KaiA, promoting dephosphorylation and turning $D$ KaiC into $S$ KaiC, leading a relatively small amount of $S$ KaiC to produce more $S$ KaiC, and to inhibit its phosphorylation into $D$ KaiC. Therefore, in the monomer model, KaiA sequestration effectively acts to catalyze the production of $S$ KaiC, which in turn causes more sequestration. This suggests that the monomer model is an example of a relaxation oscillator (Figure 3), in which the cycle can progress only when enough free KaiA is sequestered to trigger the strong positive feedback which causes the sequestration of all free KaiA, at which point the KaiC monomers can fully dephosphorylate.

The model is described by the following system of differential equations [11]:

$$\frac{dT}{dt} = k_{UT}(S)U + k_{DT}(S)D - k_{TU}(S)T - k_{TD}(S)T \tag{7}$$

$$\frac{dD}{dt} = k_{TD}(S)T + k_{SD}(S)S - k_{DT}(S)D - k_{DS}(S)D \tag{8}$$

$$\frac{dS}{dt} = k_{US}(S)U + k_{DS}(S)D - k_{SU}(S)S - k_{SD}(S)S. \tag{9}$$

The $U$ concentration is then determined by the conservation of total KaiC:

$$C_T = U(t) + S(t) + T(t) + D(t). \tag{10}$$

The amount of free KaiA is given by :

$$A(S) = \max\{0, A_T - mS\}. \tag{11}$$

Here the sequestration stoichiometry $m = 2$ by default but we will treat it as a parameter to be varied in the subsequent section. The model then in effect assumes that KaiA has infinite affinity for $S$ -KaiC.

The $S$ dependence of each of the reaction rates is given by (with $\alpha$ and $\beta$ standing in for $U$, $T$, $S$, or $D$):

$$k_{\alpha\beta}(S) = k_{\alpha\beta}^{0} + \frac{k_{\alpha\beta}^{A} A(S)}{K_{1/2} + A(S)}. \quad 12$$

Again, we will often consider a phosphorylation fraction $p(t)$. It is defined here as the sum of the concentrations of all phosphorylated forms of KaiC:

$$p(t) = T(t) + D(t) + S(t). \quad 13$$

Unless otherwise noted the parameters for this model are those found in [11] in table S2 of the supporting information.

## Two-site allosteric model

Although our main focus is comparison of simple models that are relatively pure delay or relaxation oscillators, below we also investigate whether our conclusions carry over to a more complex, biochemically realistic model. In particular, we consider the two-site allosteric model, described in detail in [1, 32]. (We use parameters taken from Tables 2 and 3 of [32] unless otherwise noted.)

Like the allosteric model treated above, the two-site allosteric model describes the Kai system at the level of individual hexamers. Contrary to the simple allosteric model, however, the two-site model also explicitly describes the state of individual monomers, and in particular their serine and threonine phosphorylation sites, as shown Figure 5. Furthermore, each monomer now has two domains called the N-terminal and C-terminal domain. KaiA can bind to the C-terminal domain, where it will enhance the phosphorylation of all the monomers in the hexamer. Each monomer in the hexamer is phosphorylated in a well-defined order: First the threonine site is phosphorylated and then the serine site. Phosphorylation of the two sites has an antagonistic effect on the conformational state of the hexamer: The $U$ and $T$ states stabilize the active conformation and the $D$ and $S$ states stabilize the inactive conformation. Due to this antagonism, the relative stability of the conformations does not depend on the absolute number of monomers in a certain state, as is the case in the allosteric model, but rather on the difference between the number of phosphorylated threonine and serine sites. Roughly, when more serine sites are phosphorylated than threonine sites, the hexamer will switch conformation. After flipping to the inactive state, the hexamer binds KaiB on its N-terminal domain. In the model, KaiA is sequestered by the N-terminal domain only after 6 KaiB monomers are bound. The resulting delay allows hexamers lagging behind the main population to continue phosphorylation and reach the inactive state, which is essential for this model to generate robust oscillations.

Since each monomer is modeled as having 4 phosphorylation states, which all play a role in determining the allosteric state of the whole hexamer, the number of states in the model is combinatorially large. Because of this, we follow the time evolution of the system using a kintetic Monte Carlo algorithm. Given this large number of states, as well as the way in which sequestration negatively feeds back on a different part of the cycle, it seems plausible that this model represents an oscillator primarily driven by delayed negative feedback, but it is more ambiguous than the fairly direct case of the simple allosteric model.

# Results

The models presented in the previous section differ in their assumptions, in particular about the type of feedback introduced by KaiA sequestration. Since enzyme sequestration has been identified as being crucial for synchronization of individual molecular oscillators into coherent population-level rhythms in the Kai system [5], it is reasonable to expect that these differences have important consequences for the dynamics of these models.

Sequestration blocks the progression of the oscillation in each of the models described here in different ways. In the allosteric model sequestration acts to keep active KaiC hexamers from beginning to phosphorylate before enough inactive hexamers have fully dephosphorylated and released their KaiA. This effectively causes KaiA sequestration to feed back negatively, with a delay, on the phosphorylation of active KaiC. On the other hand, in the monomer model, the dynamical effect of KaiA sequestration is to cause $S$ KaiC to induce its own production, leading to the full sequestration of all free KaiA before the cycle can advance. This results in strongly bistable behavior, with KaiA sequestration controlling the switch between two slowly-evolving states. This mechanism, blocking the progression of the oscillation until sufficient sequestration has occurred, is qualitatively different from that in the allosteric model, most notably in that the block is relieved by changing the concentration of free KaiA in the opposite direction.

To investigate how this fundamental difference affects the behavior of the oscillators we vary the KaiA sequestration stoichiometry $m$ (defined by equations 3 & 11, for the allosteric and monomer models, respectively), understood as a continuous variable describing the average number of KaiA monomers sequestered per KaiC in each model (per KaiC hexamer in the allosteric model and per KaiC monomer in the monomer model). We find that changing $m$ has the opposite effect on both the amplitude and the period of the oscillation in the two different types of models. Directly modifying $m$ continuously is of course only possible in abstract mathematical models, and cannot be related directly to experiment. In order to relate to realizable systems we will then show that modifying the models to explicitly include a competitive binding protein for the KaiA sequestration site on KaiC produces the same qualitative results as directly varying the sequestration stoichiometry. We show that our results extend to common variants and extensions of the basic allosteric and monomer models in the (see SM).

## Allosteric model: Less efficient sequestration decreases period

We first consider the allosteric model. Figure 6a shows three time traces of the fraction of phosphorylated KaiC $p(t)$ (defined in equation 6). These show that increasing the sequestration stoichiometry $m$ and thus the efficiency of sequestration increases both the amplitude of the oscillation and the period. Figure 6b shows that this behavior is consistently observed over a fairly broad range of values of $m$ .

To understand this observation mechanistically, consider the effect of decreasing $m$. Since $m$ only controls the sequestration of KaiA, changing it has no direct effect on the active KaiC and therefore no direct effect on the dynamics of phosphorylating KaiC. It also does not affect the dephosphorylation rates. Decreasing $m$ can only affect the timing of KaiA sequestration. If $m$ decreases, the same amount of KaiC sequesters less KaiA. Thus there is relatively more free KaiA, including during the phosphorylation portion of the cycle (Figure 6d). This has two consequences: first, more active KaiC can begin phosphorylating while a substantial amount is still in the inactive state, and the oscillations of individual KaiC hexamers are less synchronized (Figure 6c). Therefore there is more unphosphorylated KaiC when $p(t)$ reaches its maximum and less when it reaches its minimum, explaining the decrease in the amplitude of the oscillation. Second, the phosphorylation phase can begin sooner, since when an inactive hexamer releases its sequestered KaiA and becomes an active hexamer, the inactive hexamers that remain are less able to sequester the newly released KaiA. This causes more KaiA to be freed sooner, accelerating the phosphorylation phase.

## Monomer model: Less efficient sequestration increases period

The situation is reversed in the monomer model, as can be seen in Figure 7, which shows that both the amplitude and the period decrease with increasing $m$. As in the allosteric model, this behavior can be understood mechanistically. A fundamental difference between this model and the allosteric model is that in the monomer model dephosphorylation can only begin once a certain threshold amount of KaiA has been sequestered, since the balance between phosphorylation and dephosphorylation is directly dependent on the concentration of free KaiA. This, combined with positive feedback whereby sequestration favors the $D \rightarrow S$ transition, which in turn favors more sequestration, causes the model to produce switch-like behavior. In addition, this model also shows a strong separation of timescales. Once a certain amount of KaiA is sequestered dephosphorylation occurs relatively quickly (Figure 7a) compared to the time it takes to recover from dephosphorylation. Effectively, decreasing $m$ increases the amount of S-KaiC necessary to reset the switch controlling the balance between phosphorylation and dephosphorylation. This means that the duration between most KaiC becoming fully phosphorylated and the switch resetting must increase, since S-KaiC builds up very slowly until the strong positive feedback kicks in and rapidly causes the remaining KaiA to be sequestered. The duration where all KaiA is sequestered (Figure 7d ) does not change significantly with $m$; increasing $m$ instead decreases the amount of $S$ needed to fully sequester all of the KaiA. This can be usefully contrasted with Figure 6d which shows that the allosteric model does not even need to sequester all of the KaiA in order to function as an oscillator.

### Analytic perspective on changing $m$ in the monomer model

Because the monomer model is low dimensional it is possible to understand these numerical results analytically. In order to make a more direct analysis we reduced the model from a 3 dimensional system to a 2 dimensional system by assuming that the phosphorylation and dephosphorylation of the threonine residue happen fast compared to that of the serine residue. This is a reasonable assumption since one of the observations of the original model is that the $S$ phosphorylation is much slower than $T$ phosphorylation [11]. In this limit $U \leftrightharpoons T$ is in steady state. Even though $S \leftrightharpoons D$ is fast, it is bistable so it cannot be set to steady state. With this in mind we change to a new set of variables $X = S - D$ and $Y = S + D$. We then have one fast variable, $X$, which describes the resetting of the switch, and one slow variable, $Y$, which describes the of dynamics of the serine phosphorylation state. Figure 8d shows the nullclines of the resulting 2 dimensional system in terms of $X$ and $Y$. They can be seen to form the characteristic shape of a relaxation oscillator, describing slow evolution near the red nullcline and fast evolution between the two different branches of the nullcline. It is also possible to use this reduced system to provide an analytical explanation for the direction of the period dependence on $m$ (See SM).

We can use the intuition gained from this asymptotic analysis to look for traces of this behavior in the full model. For example, we can predict that if the oscillator takes longer to reach its threshold, $X$ will still only be active very briefly, since it describes the switching and is controlled by the dephosphorylation dynamics, which do not depend on $m$. Additionally, jumps in $X$ should be roughly coincident with changes in the sign of the derivative of $Y$, since the jumps between the two branches of the nullcline (indicating a large change in $X$) are the indicators that the slow dynamics of the system (indicated by $Y$) have reversed their direction. In Figure 8 we can observe the full system in terms of the variables $X$ and $Y$ exhibiting these characteristic behaviors. These behaviors appear to be crucial to the functioning of the oscillator since they are present even when the sequestration becomes very inefficient, up until the oscillation ceases. Additionally it is possible to see that the majority of the effect on the period is an increase in the amount of time with little $S$, consistent with the finding for the reduced model that the amount of time spent unsequestered does not depend on $m$.

## Competitive binding effectively modulates $m$

It is not possible to vary the parameter $m$ directly in an experiment. A direct way to emulate changing the sequestration efficiency is instead to introduce a protein that can bind competitively with KaiA in the KaiB-KaiC sequestration complex but does not promote KaiC autophosphorylation (Figure 9). This could be a truncated form of KaiA or a different protein that binds competitively with KaiA to the KaiC-KaiB complex (such as possibly CikA [28]). We will call this “decoy KaiA” (dKaiA or dA), and unless otherwise stated it binds to the KaiB-KaiC sequestration complex with equal affinity to KaiA. Both models considered can be modified to include this interaction, and we will show that this modification produces the same result as varying $m$ directly: although the maximum possible number of KaiA dimers sequestered does not change, the effective number of sites available is smaller due to some being occupied by dKaiA (Figure 9).

### Allosteric model

Here we introduce to the standard allosteric model the following interactions:

$$B_2\tilde{C}_i + m dA \leftrightharpoons dA_m B_2 \tilde{C}_i, \qquad 14$$

$$dA_m B_2 \tilde{C}_i \leftrightharpoons dA_m B_2 \tilde{C}_{i+1}. \qquad 15$$

We see in Figure 10a that the addition of dKaiA to the allosteric model, whole holding $m$ at a constant value of 2, shows the same behavior as changing $m$ directly. This indicates that dKaiA competing for sequestration with KaiA causes KaiA to become unsequestered faster. Increasing the amount of free KaiA allows those KaiC that transition from the inactive conformation to the active conformation early to begin phosphorylating sooner. This essentially decreases the effect of the delay in the system, and since this model is primarily a negative feedback-delay oscillator this also corresponds to a decrease in amplitude.

### Monomer model

As described, the monomer model foes not explicitly model formation of full KaiABC complex but instead assumes KaiA has infinite affinity for S-KaiC. If the decoy KaiA binds with equal strength to the KaiC-KaiB complex this amounts to a direct modification of $m$, where $m$ is modulated by $1/\left(1+dA_T/A_T\right)$. Figure 10 shows the amplitude and the period of the oscillation as a function of dKaiA concentration. For differing binding rates it is not as simple but Figure 17b,c (SM) shows that this behavior is not contingent on having equal binding rates. This suggests that the relaxation oscillator type of positive feedback present in the original monomer model is still operating in the same qualitative way. The same amount of KaiA must be sequesered to trigger the positive feedback on the $S$ phosphorylation and is simply sequestered more slowly in the presence of dKaiA, causing the period to increase.

This shows that these two models with opposing feedback properties can be distinguished by the introduction of a competitor for the KaiA sequestration site. Additionally, the effect of such a competitor can be arrived at by modulating the effective KaiA sequestration stoichiometry $m$. We will now observe the effects of doing so on a more complex model, the previously introduced two-domain allosteric model.

## Two-site allosteric model reproduces result of allosteric model

While the allosteric and monomer models are useful for analyzing the system by the virtue of their dynamics being transparent, it is valuable to understand how the results manifest in the more biologically realistic two-site allosteric model.

To simulate the effect of a competitor species, which competes with KaiA for free binding sites on the N-terminal domain of KaiC, we explicitly introduce a new protein in the two-site allosteric model which we assume to be dKaiA. Because the model tracks a discrete number of proteins, we cannot continuously decrease the sequestration capacity $m$ of a hexamer to simulate the effect of dKaiA, as is done in the other two models. Instead, in the two-site allosteric model, the N-terminal domain of a hexamer can maximally sequester six proteins, each of which can be either KaiA or dKaiA. dKaiA can only be sequestered from solution when six KaiB monomers are bound to the N-terminal domain of KaiC. Just like the binding of KaiA on N-terminal in this model, dKaiA does not bind cooperatively.

In Figure 11 we show a heat plot of the change in the period of the phosphorylation oscillator as a function of the dKaiA concentration and the affinity of dKaiA for KaiC. (We define the affinity, measured in $\mu M^{-1}$, of a binding reaction to be the inverse of the dissociation constant $K$.) In this plot, we only show results where dKaiA has a low affinity compared to the affinity of KaiA for N-terminal bound KaiB, which is $1/K_{\mathrm{eq}}^{\mathrm{N \cdot KaiA}} = 10^7 \mu M^{-1}$. Clearly, for all dKaiA concentrations and affinities shown in Figure 11, the period of the oscillation is less than in the absence of dKaiA. Both the dKaiA concentration we use in our simulations and the resulting decrease in period are in good quantitative agreement with the experimental results shown in [29]. Also, when we look at time traces of the phosphorylation level in Figure 12, it is clear that the troughs of the oscillation move up with increasing initial concentrations of dKaiA. This also agrees well with experiments.

Our simulations show that for the two-site allosteric model, adding a protein that competes with KaiA for the binding sites on the N-terminal domain reduces the period of the oscillator. The period reduces because, by blocking the KaiA binding sites, dKaiA decreases the the time that KaiC can sequester all KaiA from solution. This also explains why the trough of $p(t)$ increases with the dKaiA level: Due to competition, a single hexamer can on average sequester fewer KaiA dimers. Because there a now more hexamers required to sequester all KaiA, and because these hexamers have a higher phosphorylation level compared to those that have already flipped back to the active state, the trough in the phosphorylation level moves up. The fact that the period decreases shows that, just as we concluded for the allosteric model, the oscillator in the two-site allosteric model behaves as a delay oscillator.

Because, in these calculations, dKaiA has a much lower affinity than KaiA for the N-terminal domain of KaiC, dKaiA is most effective competing with KaiA for free binding spots when there are only a few KaiA dimers in the solution. This is because the probabilities that dKaiA or KaiA will bind to the N-terminal domain are roughly proportional to $dA/K_{eq}^{N \cdot dKaiA}$ and $A/K_{eq}^{N \cdot KaiA}$, respectively, where $K_{eq}^{N \cdot \circ}$ is the dissociation constant of the associated reaction. Given that $K_{eq}^{N \cdot dKaiA} \gg K_{eq}^{N \cdot KaiA}$ in our simulations, dKaiA only has a reasonable chance to bind when the concentration of free KaiA is extremely low. This is only the case when all KaiA is sequestered by KaiC and only one or two KaiA are free in solution due to hexamers flipping back to the active states prematurely. Therefore dKaiA only has an effect at the end of the oscillation and not in the phase when most hexamers are in the active state and there is a lot of free KaiA in solution.

If, as proposed above, the two-site allosteric model can be understood as an example of a negative feedback model, these results are consistent with the simpler allosteric model. This further supports the conjecture that the effect on the period of adding dKaiA to the oscillator would be a reliable readout of the sign of the feedback present in the system.

# Discussion

A major outstanding problem in chronobiology is to discern how a clock's feedback architecture affects its properties, and thus what specific advantages different architectures may confer. A necessary initial step in this direction is to have a clear understanding of what feedback structures are actually prevalent in natural clocks. Here, we have addressed this question for the post-translational Kai protein oscillator derived from the cyanobacterium *S. elongatus*. Specifically, we have used mathematical and computational modeling to clarify the difference between two simple classes of models of the Kai system.

These models, the allosteric model and the monomer model, assume very similar molecular interactions; in particular, they both rely on KaiA sequestration to synchronize the phosphorylation states of KaiC molecules. Nonetheless, they fall into the distinct dynamical classes of delay and relaxation oscillators, respectively, depending on the effect produced by KaiA sequestration: Sequestration effectively introduces a positive feedback loop in the monomer but not in the allosteric model. We showed here that these differences cause the period of the oscillator to change in opposing directions when the KaiA sequestration stoichiometry $m$ is reduced. In the allosteric model, at lower $m$ KaiC hexamers can begin to phosphorylate earlier and the period decreases. In the monomer model, on the other hand, at lower $m$ the system needs to wait longer for enough KaiA to be sequestered to reset the balance between phosphorylation and dephosphorylation, and both the amplitude and period increase. These trends persist in extensions of the basic allosteric and monomer models discussed in the SM and in the more biochemically detailed two state allosteric model, which behaves essentially as a delay oscillator.

These different oscillation mechanisms have heretofore proven to be difficult to distinguish experimentally. In the monomer model, positive feedback arises because $S$ KaiC promotes its own production by sequestering KaiA, which in turn increases the $D$ to $S$ transition rate while decreasing the backwards $S$ to $D$ rate. In contrast, in the allosteric model KaiA is sequestered by inactive KaiC, but free KaiA has little effect either on the flip rates between the active and inactive conformations or on the transition rates between different inactive phosphoforms; such positive feedback is thus absent. The essential question is hence whether free KaiA limits the formation of species competent to sequester KaiC, and in particular, how strongly KaiA modulates the transition rates between the $D$ and $S$ forms of KaiC.

The experimental evidence on this point is mixed. On one hand, studies with phosphomimetics indicate that KaiC with a constitutively phosphorylated serine residue phosphorylates only very slowly on the threonine residue, even in the presence of free KaiA. [15, 27]. Similarly, the rate of threonine dephosphorylation appears to be independent of KaiA and KaiB [15]. These results imply that KaiA sequestration does not promote the creation of further species that sequester KaiA to a meaningful degree, that is, that KaiA sequestration does not feed back positively on itself. On the other hand, a study of the individual phosphostate time traces in the native protein found that they could only be well fit with a monomer model when the transition rates between $D$ and $S$ depended strongly on KaiA [11].

The observation that the oscillator period varies in the opposite way with $m$ for delay and relaxation oscillators provides a robust way to sidestep this ambiguous molecular evidence and to determine directly which dynamical mechanism better describes the Kai system. We showed that, in practice, $m$ can be changed by introducing an additional protein, which we refer to as decoy KaiA (dKaiA), that competes with KaiA for the binding sites on the KaiBC complex where KaiA is sequestered but does not in any way affect KaiC phosphorylation. One obvious way to construct dKaiA would be to use a truncated or mutant form of the KaiA protein. Another strong candidate for such an assay is CikA, which is a known element of the clock output mechanism and binds to the KaiBC complex. Adding CikA to the *in vitro* Kai system has been shown to decrease the oscillation amplitude and period [29]. CikA is also known to bind to the KaiA sequestration binding site [28]. Together, these findings strongly suggest that the post-translational Kai clock functions essentially as a delay oscillator.

Although descriptions based on KaiA sequestration have recently predominated, it is worth noting that there are also models of the Kai system whose operation cannot be mapped onto the two types of model considered here in an obvious way. In particular, exchange of monomers between KaiC hexamers is known to occur and has been shown in principle to contribute to the synchronization of their phosphorylation cycles [33, 34]. The observation that oscillations are lost as the KaiA concentration is increased [11, 35], however, strongly suggests that KaiA sequestration is the dominant synchronization mechanism [24].

Our focus on modeling KaiA sequestration proved to be a useful starting point for revealing crucial mechanistic details about the system. By comparing two simple but qualitatively distinct numerical models of the oscillator we could understand current experimental results in a new light, helping to contextualize relatively subtle differences that would have been difficult to interpret otherwise. Our results demonstrate that introducing competitors for important binding sites can be a useful tool to probe parameters and feedbacks in biological networks that might otherwise be difficult to study experimentally.

More broadly, our conclusions shed new light on the architecture of the post-translational Kai oscillator, a widely-used model system for biological clocks. The identification of the Kai oscillator as one driven primarily by negative feedback with a delay is also notable for what it suggests about the design principles underlying biological oscillators. Indeed, both positive [4] and negative [6] feedback loops have been proposed to make distinct contributions to oscillator robustness, and it remains unclear how selection balances these advantages to arrive at the structures of real clock networks. In the case of the Kai system, it is notable that KaiC hexamers have a large number of internal states, facilitating the introduction of delay into the negative feedback loop [22-24]. This suggests a potential parameter regime in which negative feedback oscillators could preferentially be found in biological systems.

# Acknowledgements

We are grateful to Danny Forger, Yining Lu, and Pieter Rein ten Wolde for helpful conversations. This work was funded in part by NSF grant DMR-1056456 and by the Margaret and Herman Sokol Faculty Awards.

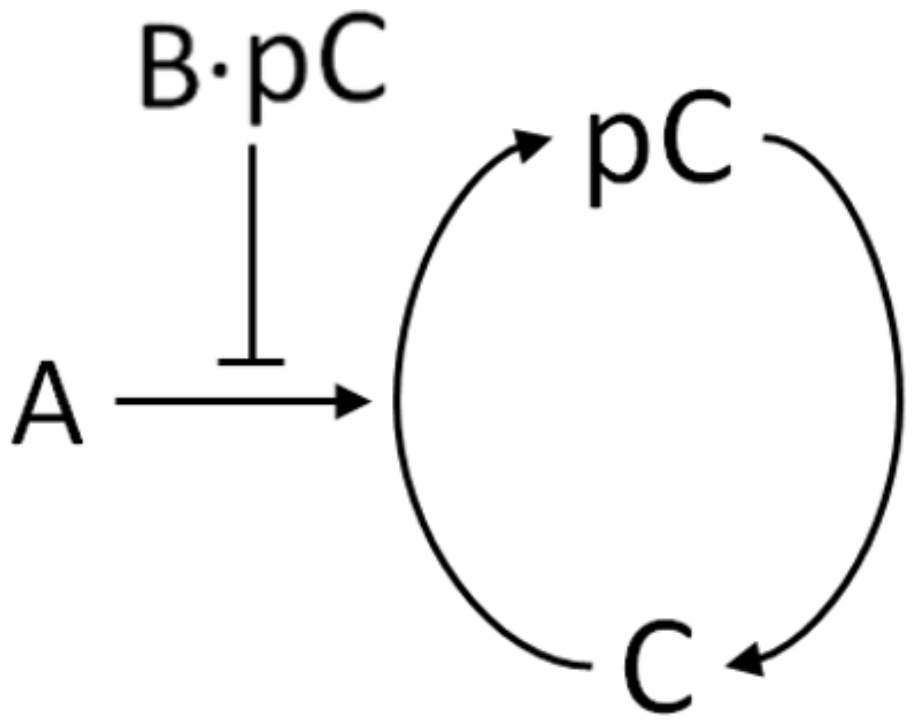

Figure 1

Schematic of the Kai oscillator. KaiC hexamers autophosphorylate on serine and threonine residues in the presence of KaiA. Phosphorylated KaiC can sequester KaiA in a KaiABC complex, causing KaiC to dephosphorylate.

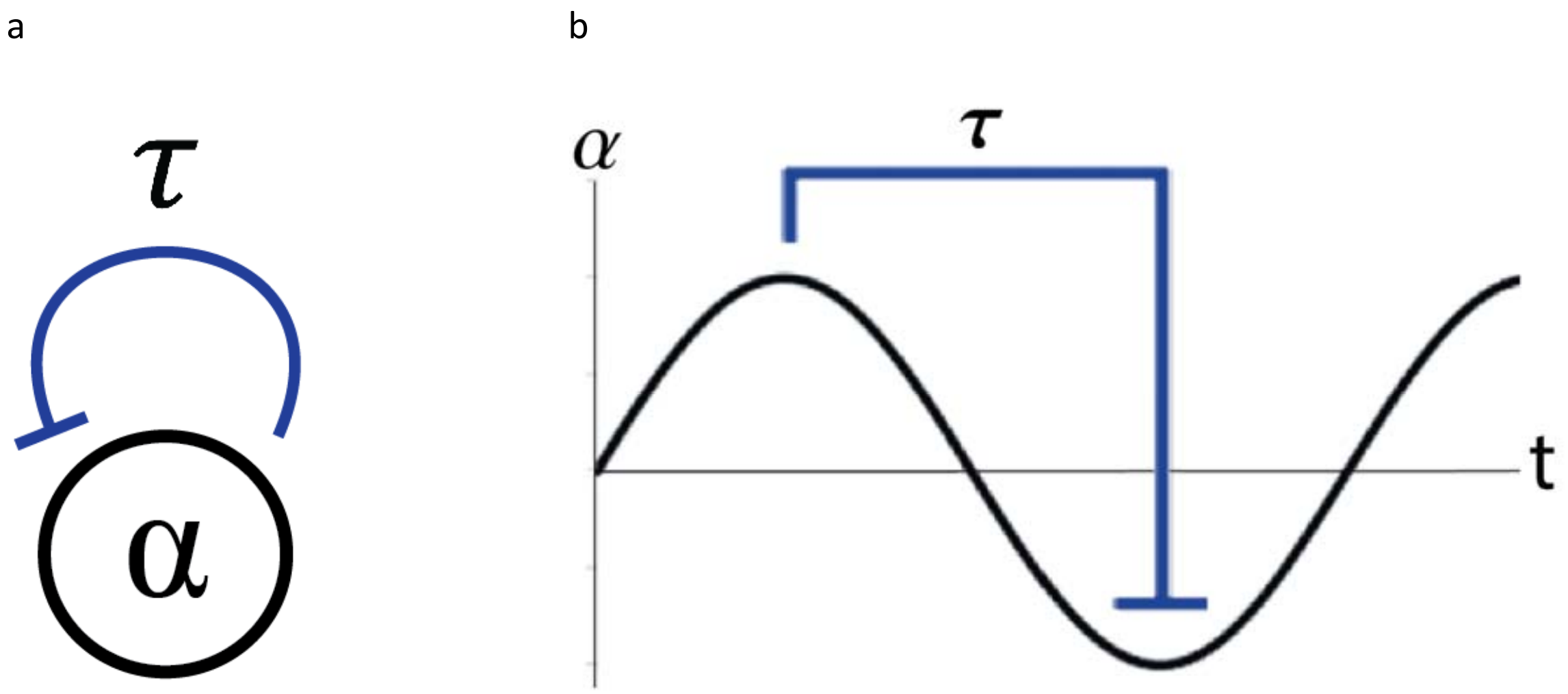

Figure 2

Delay oscillators operate by delaying negative feedback, indicated by flat arrows in blue. (a) Schematic delay oscillator: a protein $\alpha$ feeds back negatively on its own activity with time delay $\tau$ . (b) A peak in $\alpha$ activity leads to delayed inhibition, inducing a trough at a time $\tau$ later. This trough relieves the inhibition, allowing the signal to recover towards a peak after another delay of order $\tau$ .

a b

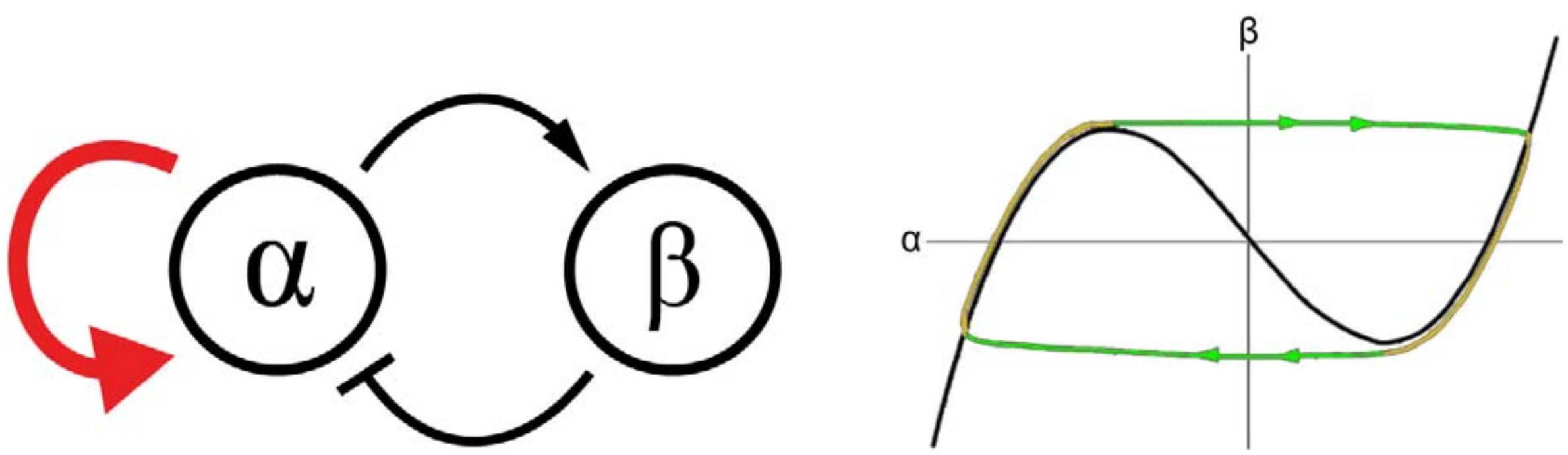

Figure 3

A relaxation oscillator operates with a combination of positive and negative feedback. (a) The simplest kind involves two species, α and β. Positive feedback, indicated with a red arrow, on one of these (α in this case) causes the system to push past its steady state instead of settling down, producing an oscillation. This is most effective when there is a strong separation of timescales, with α evolving much faster than β. (b) The relaxation oscillator limit cycle in the α-β phase plane. Positive feedback on α gives it a bistable nullcline (black). The system evolves slowly along one branch of the nullcline until one of the extrema are reached (yellow). It then moves quickly to the other branch (green) and repeats the process.

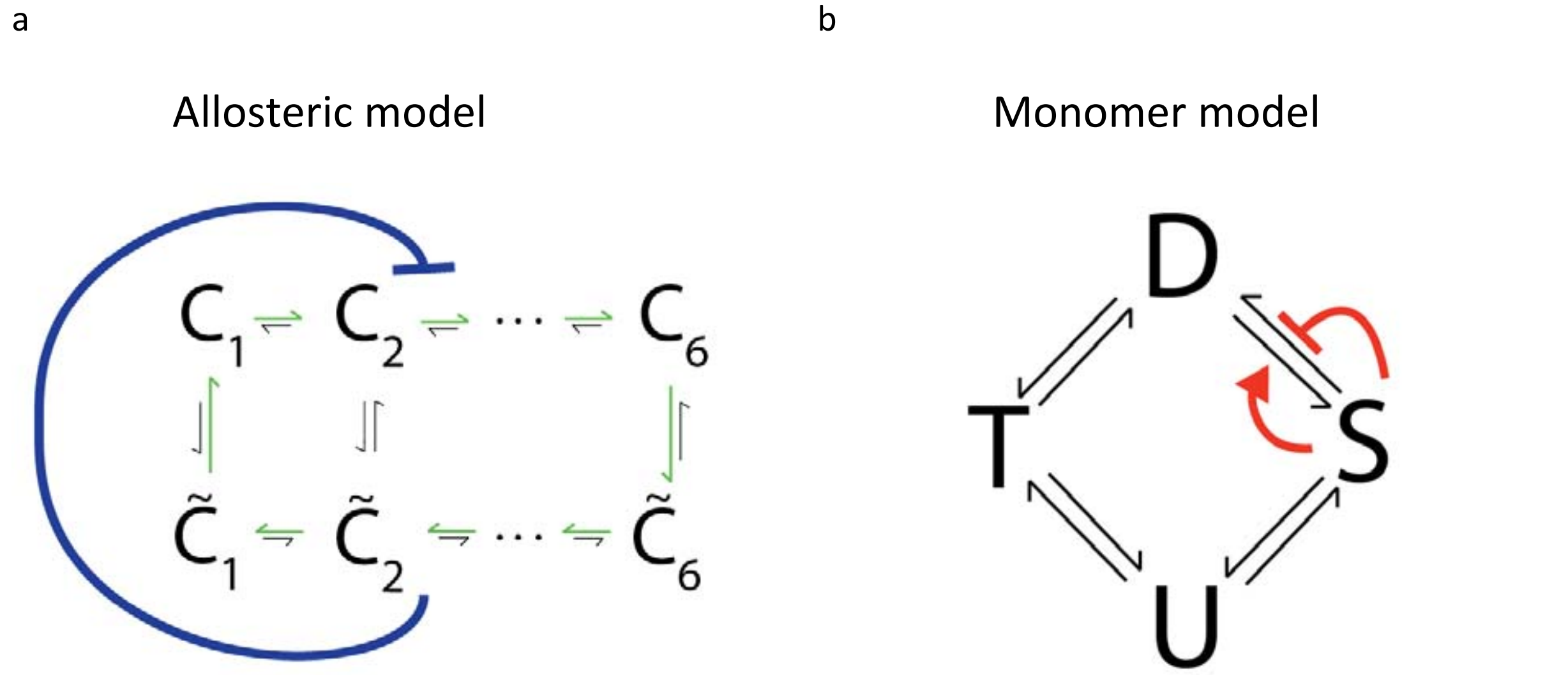

Figure 4

The allosteric and monomer models. Effective negative feedback, blue; effective positive feedback, red.

(a) In the allosteric model, the oscillation proceeds by the phosphorylation of KaiC hexamers in the active conformation, denoted $C_i$, followed by a conformational change and dephosphorylation of inactive hexamers ( $\tilde{C}_i$ ). Green arrow indicates dominant direction of conformal transitions. The inactive KaiC sequesters KaiA, preventing the active KaiC from phosphorylating, introducing the negative feedback and delay shown in the blue.

(b) The monomer model involves transitions between 4 different phosphorylation states on a KaiC monomer, where a serine and a threonine residue can each be either phosphorylated or unphosphorylated. Unphosphorylated KaiC ( $U$ ) becomes phosphorylated on the threonine ( $T$ ) and then on the serine, making it doubly phosphorylated ( $D$ ). When the threonine dephosphorylates, only the serine remains phosphorylated ( $S$ ). It is this state that can sequester KaiA. This sequestration has the effect of increasing the rate of $D \rightarrow S$ transitions while decreasing the $S \rightarrow D$ rate. Both of these interactions amount to positive feedback of $S$ on its own concentration and so are shown in red above.

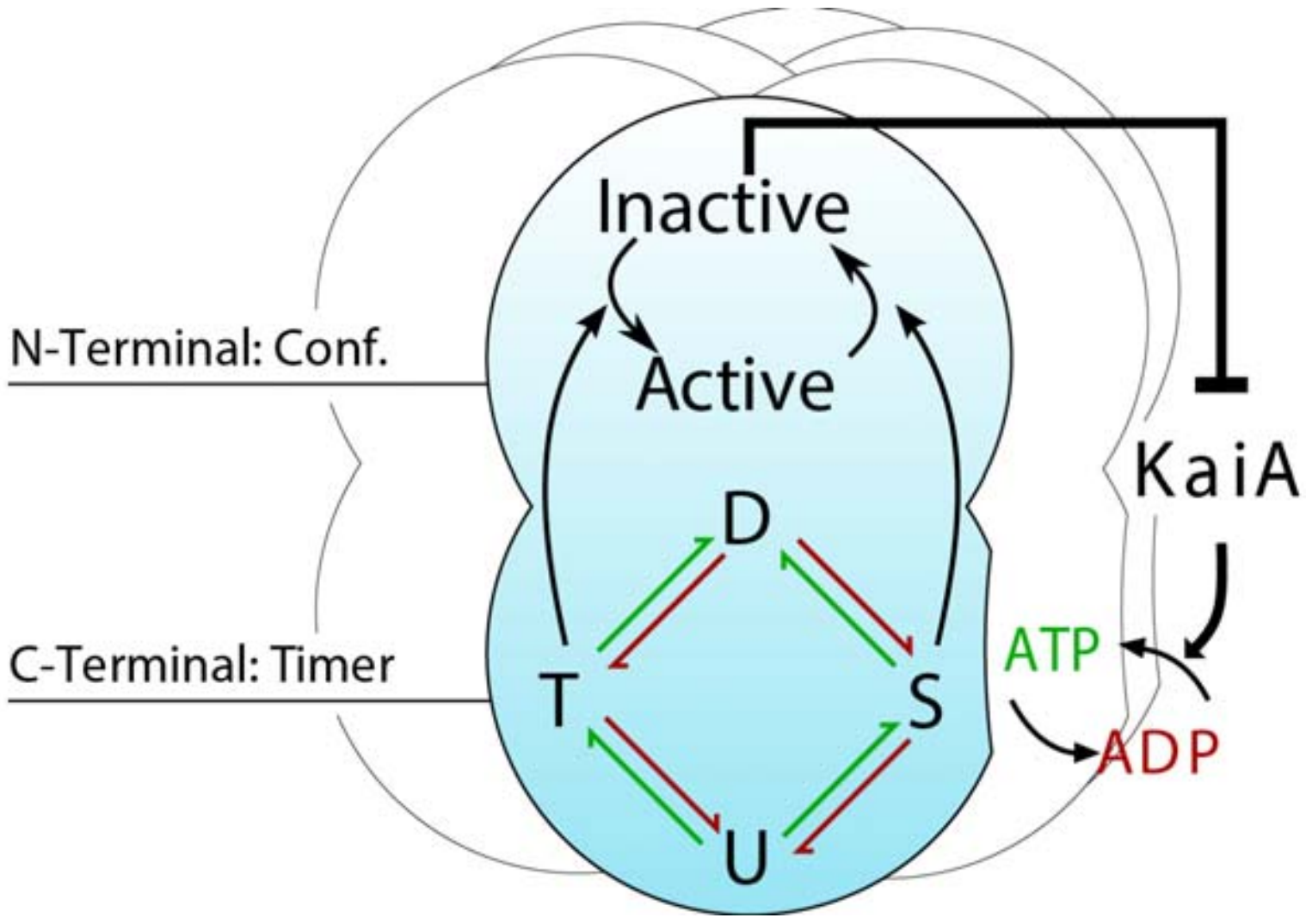

Figure 5

Cartoon of the two-site allosteric model, which accounts for both the hexameric nature of KaiC and the presence of 2 distinct phosphorylation sites each monomer. Each KaiC monomer consists of a N-terminal and a C-terminal domain: the N-terminal domain drives the conformational change of the hexamer between the active and inactive state and the C-terminal domain determines the timing of this switch through the ordered phosphorylation of its two phosphorylation sites. Phosphorylation of the threonine site (T-state) stabilizes the active state and phosphorylation of the serine site (S-state) stabilizes the inactive state. Phosphorylation can only occur with ATP (green arrows) and dephosphorylation only with ADP (red arrows) in the C-terminal binding pocket. (Figure adapted from [1].)

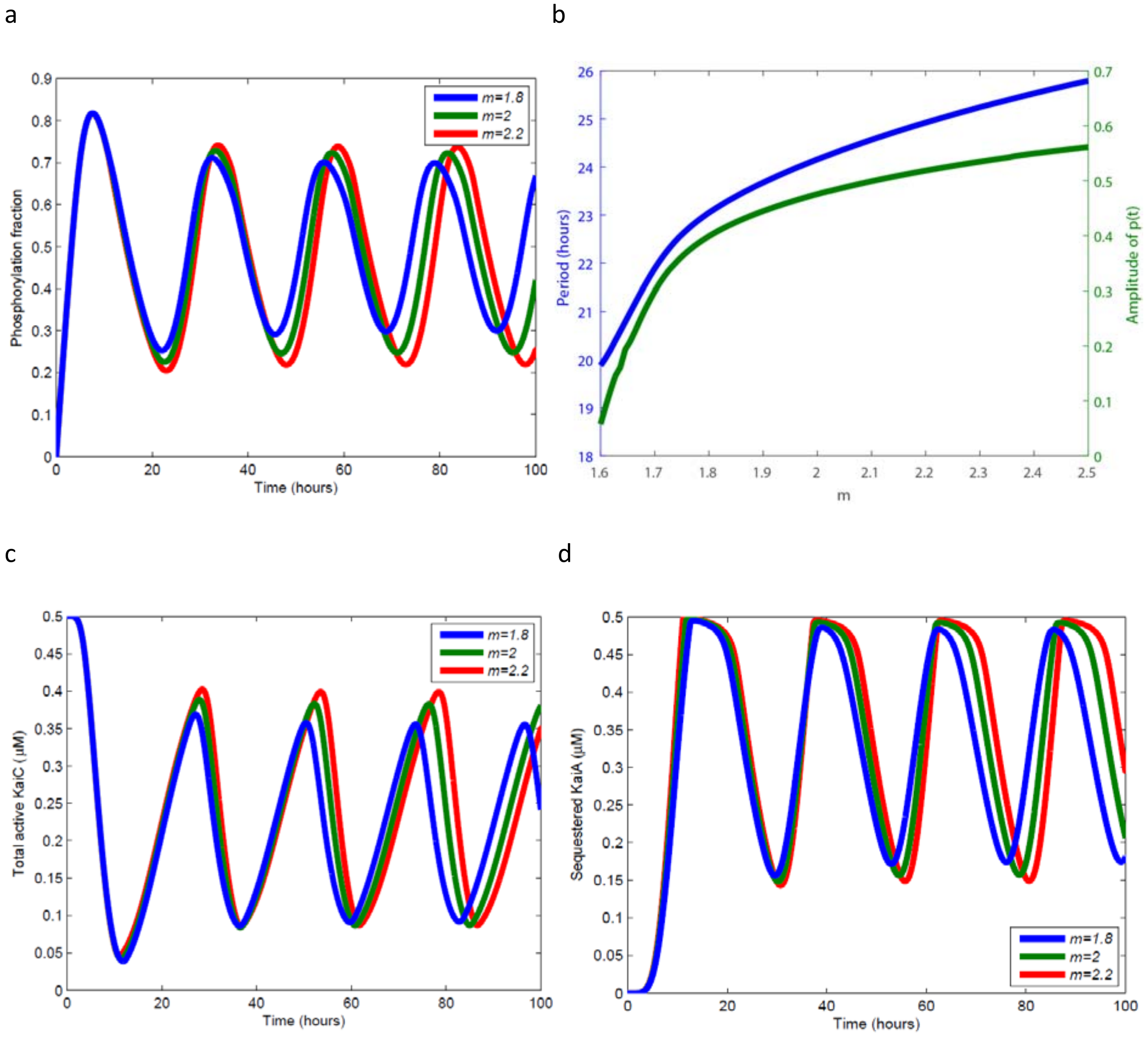

Figure 6

Varying the efficiency $m$ of KaiA sequestration in the allosteric model. (a) Time traces of the phosphorylation fraction $p(t)$ for three different values of $m$. In this model $m$ varies in the same direction as the amplitude and the period of the oscillation. (b) This relationship can be seen over a wide range of $m$. (c) The concentration of total active KaiC, $\sum_{i=0}^{6} C_i^T$, for three different values of $m$. The decrease of the maximum with $m$ indicates that at higher $m$ less KaiC has switched to the active conformation before the majority switches to the inactive conformation and begins dephosphorylating, indicating decreased synchronization. (d) The concentration of sequestered KaiA, defined as $m\sum_{i=0}^{6}\frac{A^m B_2 \tilde{C}_i^T}{\tilde{K}_i^m + A^m}$. As $m$ decreases, the maximum amount of sequestration also declines, allowing more KaiC to phosphorylate even when most KaiC is inactive. Additionally, the minimum amount of sequestered KaiA is higher, reflecting the presence of more inactive KaiC when the majority is active.

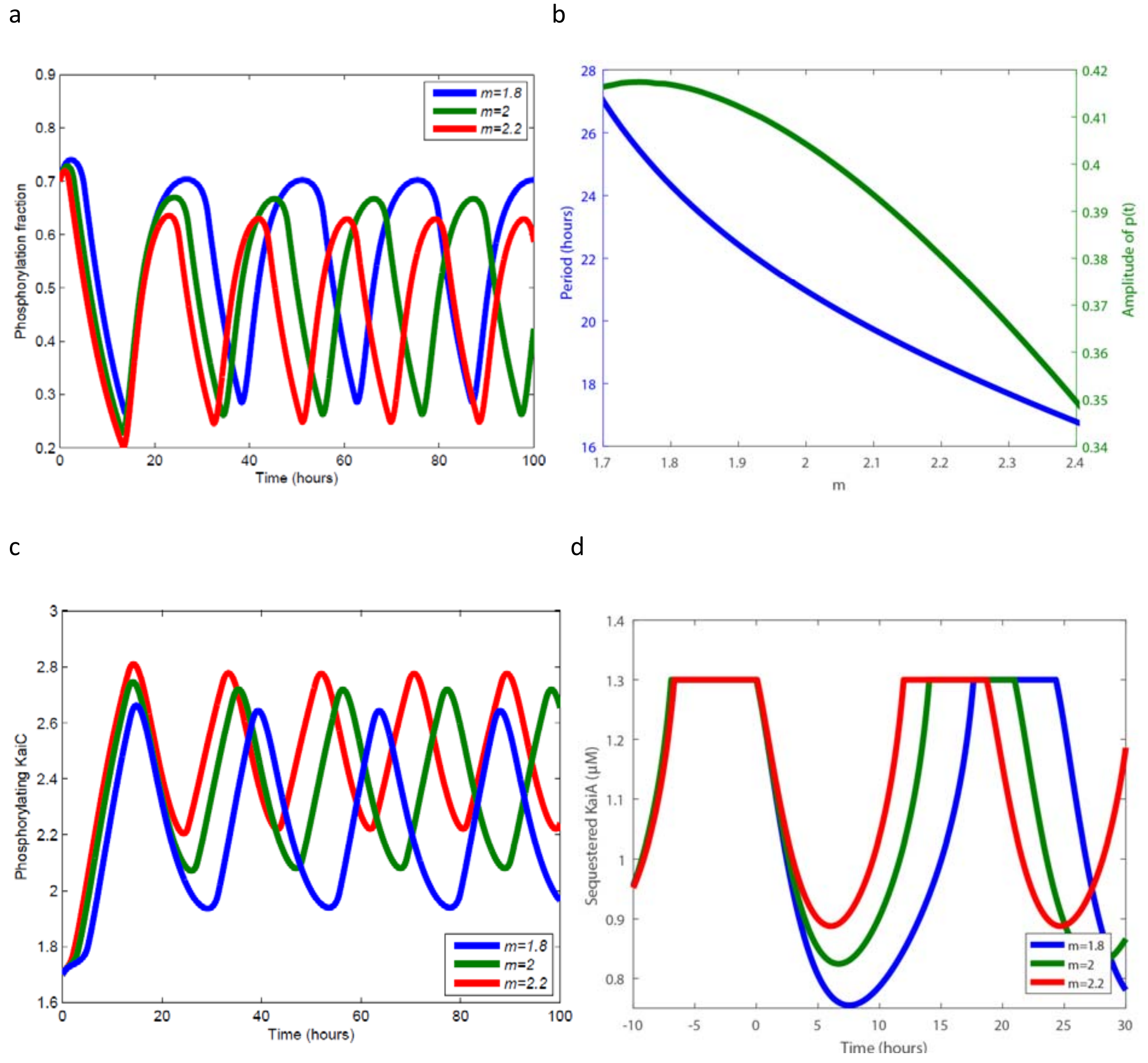

Figure 7

Varying the sequestration efficiency $m$ in the monomer model. (a) Time traces of the phosphorylation fraction $p(t)$ for three different values of $m$. In this model $m$ varies inversely with the amplitude and the period of the oscillation. (b) This relationship can be seen over a wide range of $m$. (c) The concentration of $U$ and $T$ KaiC, the closest analog to the active KaiC in the allosteric model, since these are the KaiC species that are competent to phosphorylate. Compared with Fig. 6c it can be seen that this model does not exhibit the same desynchronization as $m$ is decreased. (d) The concentration of sequestered KaiA. Traces for different values of $m$ are aligned so that dephosphorylation ends at $t = 0$. Compared to the allosteric model, this model sequesters all of the KaiA for part of the cycle, almost regardless of $m$. It can be seen that $m$ does not affect the duration of dephosphorylation does not vary strongly with $m$, and that the majority of the variation in period comes from the approach to full sequestration. Once the necessary threshold of sequestered KaiA is crossed, positive feedback on $S$ KaiC causes all of the KaiA to become sequestered.

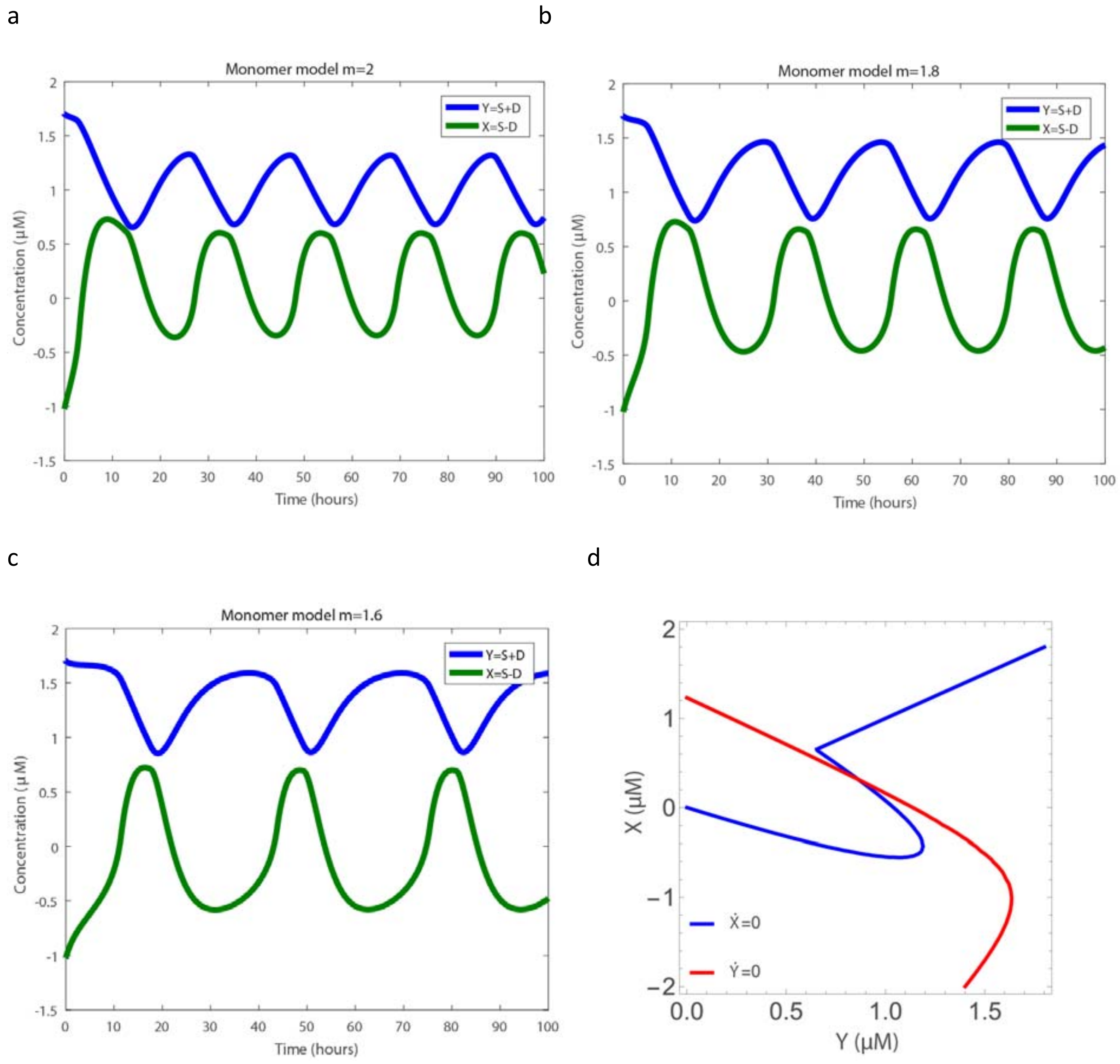

Figure 8

(a – d): Dynamics inferred from the reduced model are apparent in time traces from the full monomer model, for $m = 2$ (a), $m = 1.8$ (b), and $m = 1.6$ (c). Changes in the sign of $X = S - D$ are correlated with changes in the sign of the derivative of $Y = S + D$, indicating a switch along a fast degree of freedom. (d): A phase plane plot of the nullclines for a reduced monomer mode. The red nullcline corresponds to $X = S - D$ and the blue to $Y = S + D$. The $X$ nullcline has 3 main branches, and the middle branch is intersected by the $Y$ nullcline, a motif indicative of a relaxation oscillator. See Supporting Material for expressions for $\dot{X}$ and $\dot{Y}$.

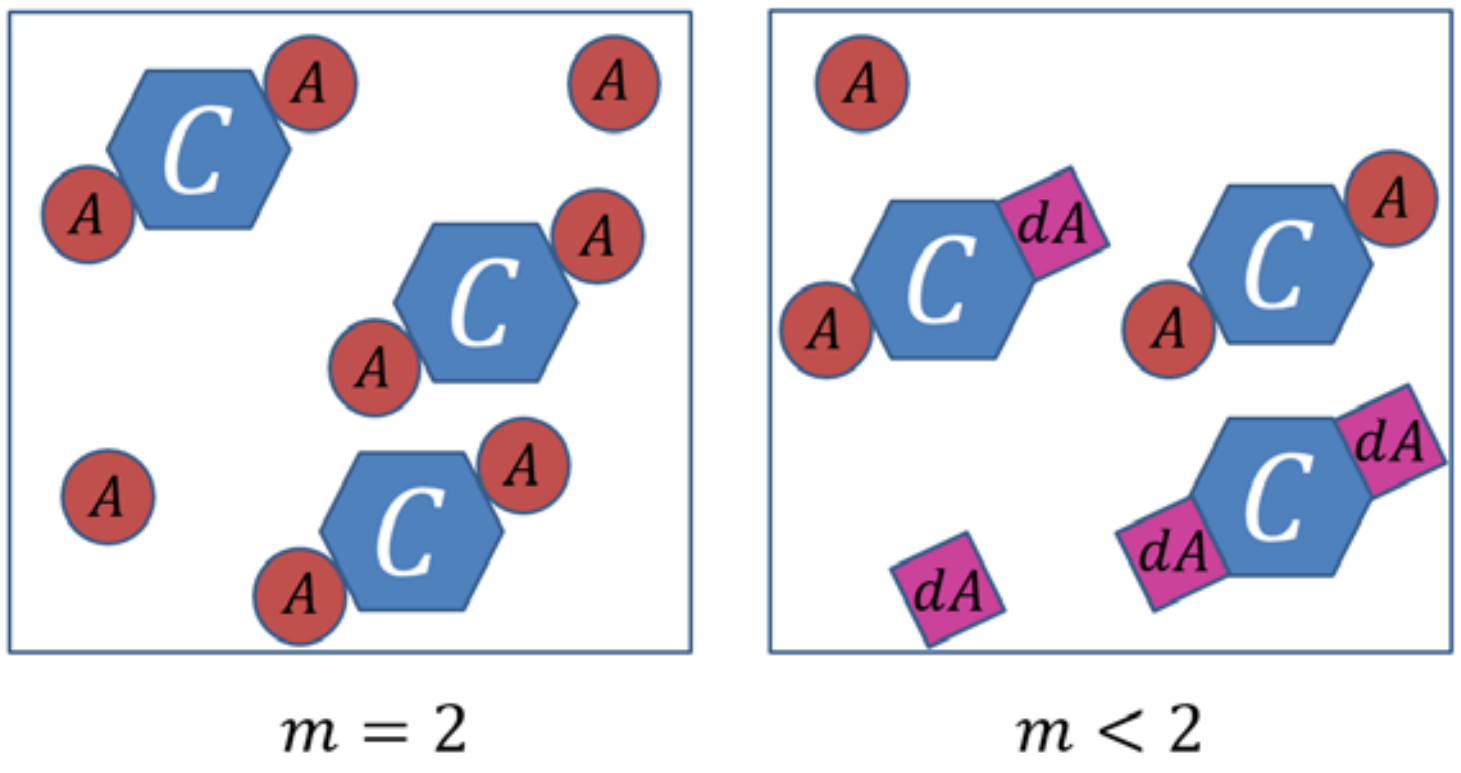

Figure 9

A schematic of a method for varying the KaiA sequestration efficiency $m$. By introducing a species (here shown as “dA” for “decoy KaiA”) that binds competitively with KaiA for the site at which KaiA is sequestered it is possible to experimentally vary the average amount of KaiA sequestered per KaiC.

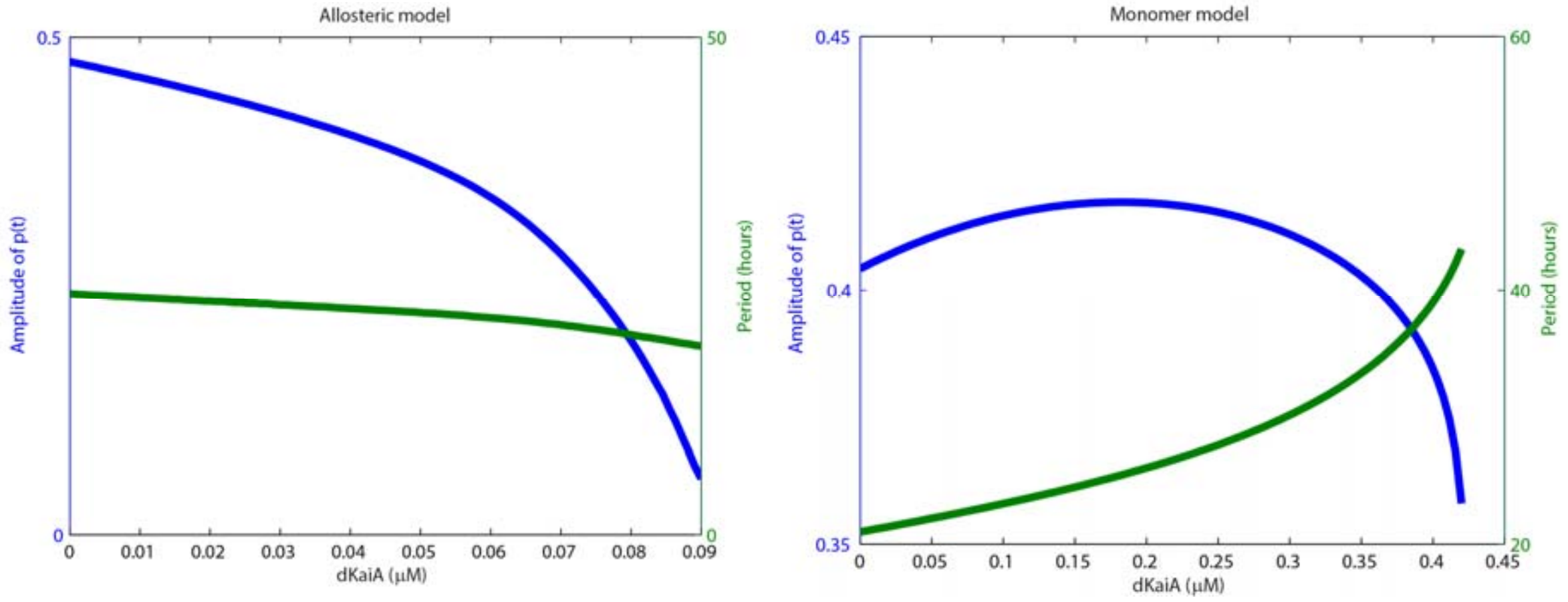

Figure 10

a) The period and amplitude of the allosteric model with the KaiA sequestration stoichiometry set to 2, but including some concentration of a decoy KaiA (“dKaiA”) which is sequestered competitively with KaiA but does not bind to KaiC otherwise (for example to promote KaiC autophosphorylation). This has qualitatively the same effect on the amplitude and period as changing $m$ directly does. Since increasing $m$ and increasing the concentration of the decoy should have opposite effects on the amount of KaiA bound, this figure shows the same qualitative behavior as Fig. 8B.

b) Amplitude and period for the original monomer model with dKaiA. Since the decoy KaiA is sequestered with the same affinity as KaiA, $m$ is simply modulated from its default value of 2 by a factor of $1/\left(1+\frac{dKaiA_T}{KaiA_T}\right)$, and is formally equivalent to changing $m$ directly.

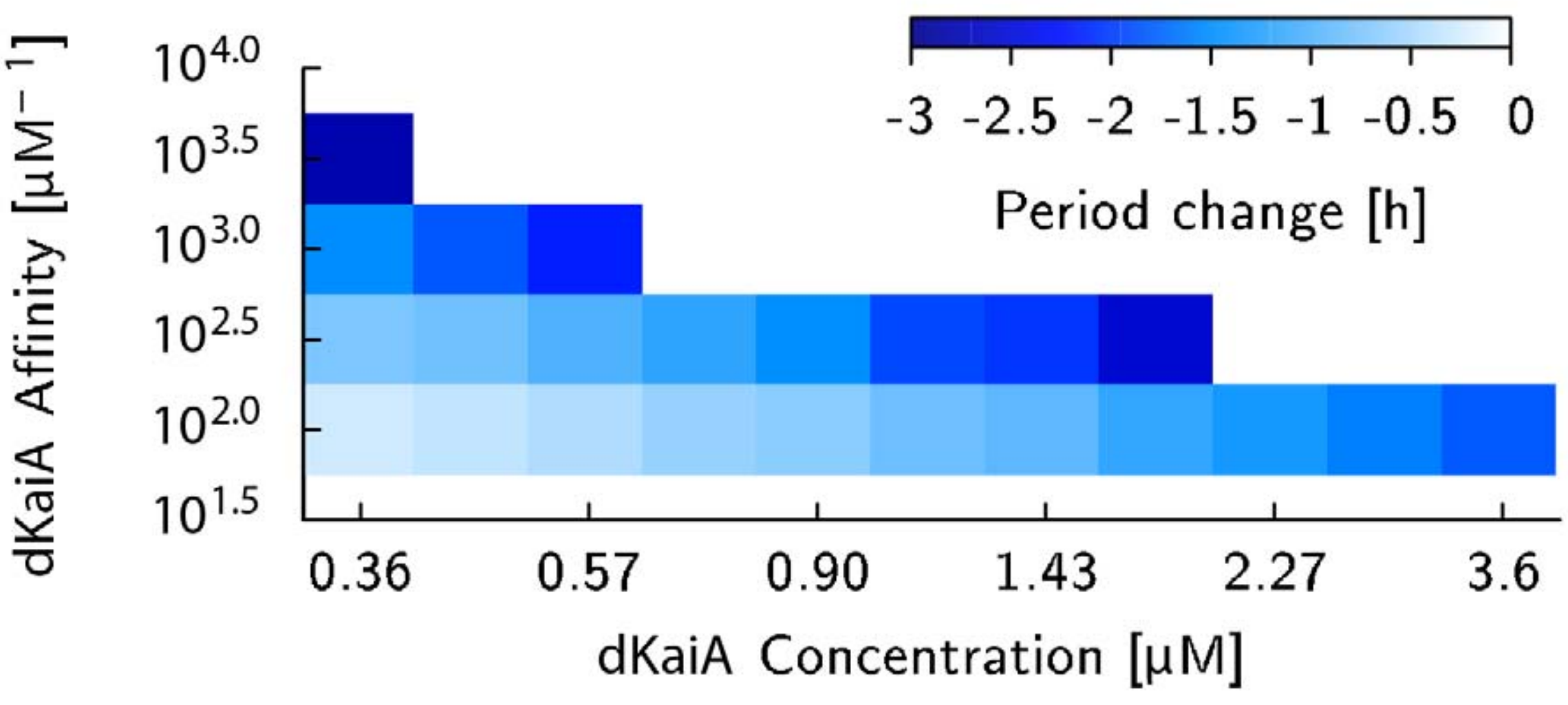

Figure 11

Heat plot showing the change in the period for the two site allosteric model as a function of the initial dKaiA concentration and the affinity of dKaiA for KaiB bound to KaiC. Note that both axes are log scale. When the dKaiA affinity is significantly lower than the KaiA affinity, which is the case for the values shown, dKaiA decreases the period of the oscillation. Parameter values corresponding to the area in white show no oscillations.

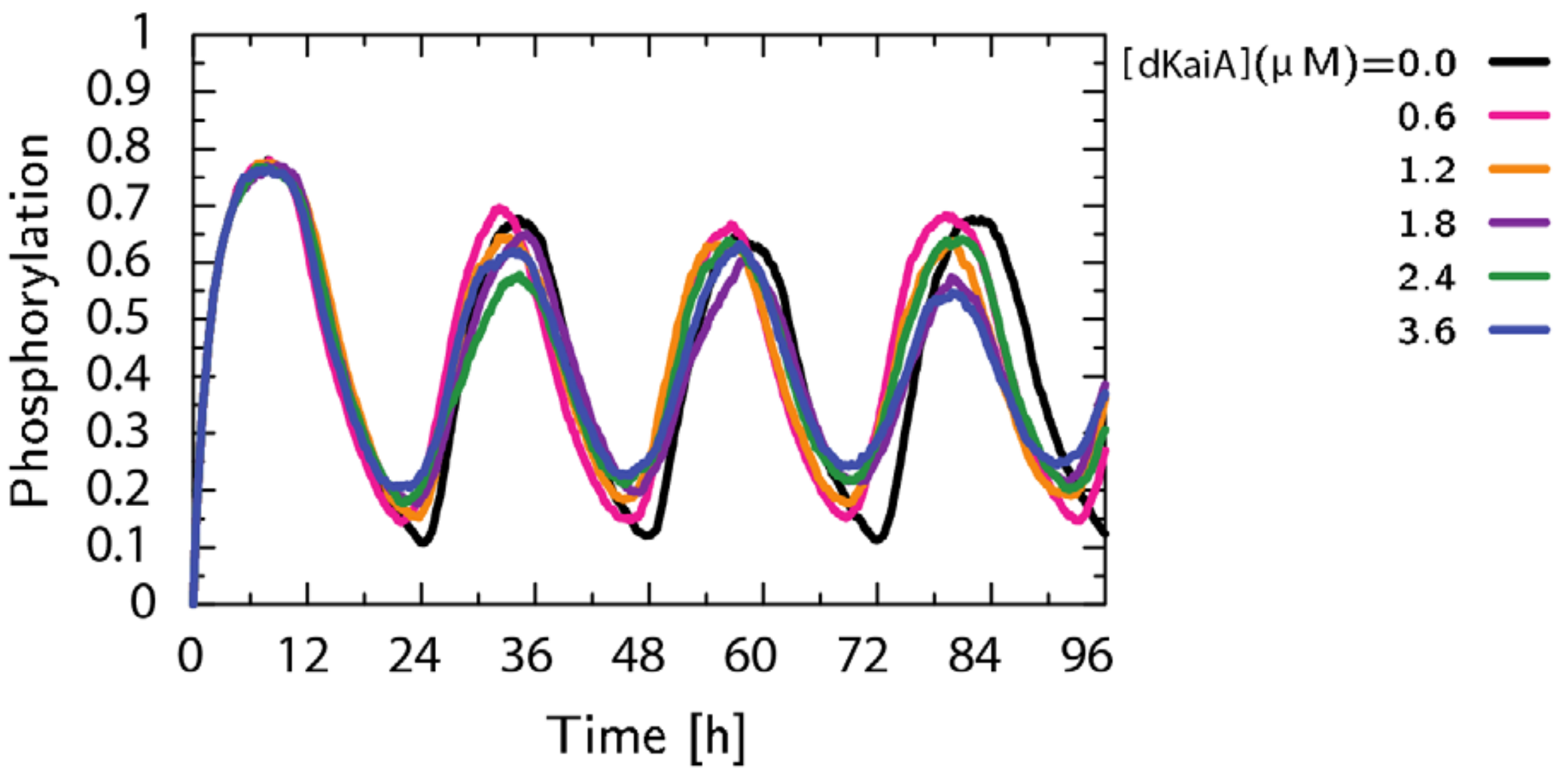

Figure 12

Time traces of the phosphorylation level $p(t)$ at different concentrations of dKaiA, for the two site allosteric model. The affinity of dKaiA for the N-terminal domain of KaiC is $K_{\mathrm{eq}}^{\mathrm{CI\cdot KaiA}}$ = 100 $\mu M$ . Consistent with experiments, the period decreases and the troughs of the oscillations move to a higher phosphorylation level with increasing dKaiA concentration.

# Supporting Material

## Modifications of basic models don't change qualitative behavior

To test how robust these results are to complications to the model structure we investigated published extensions to the previously presented models. These extensions show very similar responses to changes in sequestration stoichiometry to the original models, supporting the hypothesis that these responses represent a general distinction between dynamics that are driven primarily by delay or primarily by positive feedback.

### Allosteric model

In 2010 a number of extensions to the allosteric model were introduced [36], but the one that is most relevant here is one that allows KaiA to bind to KaiC and promote autophosphorylation in either one of the conformational states hypothesized by the allosteric model. It allows KaiA to promote autophosphorylation in the "inactive" allosteric state, adding the following interactions:

$$\tilde{C}_i + A \leftrightharpoons A\tilde{C}_i \rightarrow C_{i+1} + A, \qquad 16$$

$$A_m B_2 \tilde{C}_i + A \leftrightharpoons A_{m+1} B_2 \tilde{C}_i \rightarrow A_m B_2 C_{i+1} + A. \qquad 17$$

This introduces an element of positive feedback into the model. We assume that the affinity for KaiA to phosphorylate the "inactive" state KaiC hexamers is smaller than the affinity for "active" state KaiC hexamers by a factor of 100. This preserves the qualitative form of the dependence of the amplitude and period on $m$ as shown in Figure 13. Even though the affinity of KaiA to promote autophosphorylation in the "inactive" allosteric state is weaker than for the "active" allosteric state this model still matches the experimentally observed phenomenon of the phosphorylation fraction increasing upon adding a large amount of KaiA even when the phosphorylation fraction was decreasing, which the previous model was not able to reproduce. This effect is shown in Figure 14.

### Monomer model

The monomer model was also extended [37], in this case to include explicit KaiB binding, and allows the KaiB-bound doubly phosphorylated state to weakly sequester KaiA in addition to the state that is phosphorylated only on the serine residue. It still relies on positive feedback on the $S$ phosphorylation state, albeit now bound to KaiB, so it should exhibit a similar response to the original model to the modification of sequestration stoichiometry. We see this is supported in Figure 15. This model differs from the 2007 version of the model by adding an irreversible step that corresponds to ATPase activity allowing KaiB to bind to the $S$ and $D$ states. These KaiB bound monomers are then the ones that participate in sequestering KaiA. $D$ does in fact participate in KaiA sequestration, but only to 2% of the extent to which $S$ sequesters KaiA, according to the published parameter set. The ODEs that govern the system are then:

$$\frac{dU}{dt} = k_{TU}(S)T + k_{SU}(S)S + k_{SBU}(S)SB - k_{UT}(S)U - k_{US}(S)U, \qquad 18$$

$$\frac{dS}{dt} = k_{US}\left(S\right)U + k_{DS}\left(S\right)D - k_{SU}\left(S\right)S - k_{SD}\left(S\right)S - k_{bc}S, \tag{19}$$

$$\frac{dT}{dt} = k_{UT}\left(S\right)U + k_{DT}\left(S\right)D - k_{DBT}\left(S\right)DB - k_{TU}\left(S\right)T - k_{TD}\left(S\right)T, \tag{20}$$

$$\frac{dD}{dt} = k_{TD}\left(S\right)T + k_{SD}\left(S\right)S - k_{DT}\left(S\right)D - k_{DS}\left(S\right)D - k_{bc}D, \tag{21}$$

$$\frac{dDB}{dt} = k_{bc}D + k_{SBDB}\left(S\right)SB - k_{DBSB}\left(S\right)DB - k_{DBT}\left(S\right)DB, \tag{22}$$

$$\frac{dSB}{dt} = k_{bc}S + k_{DBSB}\left(S\right)DB - k_{SBDB}\left(S\right)SB - k_{SBU}\left(S\right)SB. \tag{23}$$

Where the $S$ dependence of the reaction rates is the same as before, and $k_{bc}$ is an $S$ independent rate of ATPase triggered catalysis of complex formation. The amount of free KaiA is given by $A = \max\{0, A_T - mSB - nDB\}$. Unless otherwise stated the parameters are those given by table S5 in [37].

By comparing Figure 16 to Figure 8 it is possible to see that it shares the key characteristics that indicate that the dynamics described in the reduced model are still dominant for this extension. Specifically, changes in the sign of the derivative of $Y$ are associated with large changes in the magnitude of $X$, and the majority of the effect on the period is from the time when most of the KaiA is unsequestered. This suggests that this behavior is generic or at least very common in models that involve strong positive feedback as the primary driver of the oscillation. Here the effective sequestration stoichiometry for $D$ $DB$ is changed in proportion with that for $SB$.

Thus the effect of varying the stoichiometry $m$ of KaiA sequestration is robust to minor changes in the models studied here. This suggests that the effect of introducing a competitive binder for the KaiA sequestration site on the amplitude and the period is a reliable indicator of the sign of the feedback that KaiA sequestration introduces into the dynamics of the *S. Elongatus* circadian system.

## ODEs for Allosteric Model

The allosteric model is described by the following differential equations:

$$\frac{dC_{i,T}}{dt} = \sigma_{i-1}^{ps}C_{i-1,T} + \sigma_{i+1}^{dps}C_{i,T} - (\sigma_i^{ps} + \sigma_i^{dps})C_{i,T} - \sigma_i^{Ff}C_{i,T} + \sigma_i^{Fb}\tilde{C}_i, \tag{24}$$

$$\frac{d\tilde{C}_i}{dt} = \tilde{k}_{ps}\tilde{C}_{i-1} + \tilde{k}_{dps}\tilde{C}_{i+1} - (\tilde{k}_{ps} + \tilde{k}_{dps})\tilde{C}_i + \sigma_i^{Ff} C_{i,T} - \sigma_i^{Fb}\tilde{C}_i$$

$$-k_i^{Bf}\left(B_T - 2\sum_i [B_2\tilde{C}_i]_T\right)^2 \tilde{C}_i + \frac{k_i^{Bb}\tilde{K}_i\,[B_2\tilde{C}_i]_T}{\tilde{K}_i + A^m}, \quad 25$$

$$\frac{d[B_2\tilde{C}_i]_T}{dt} = \tilde{k}_{ps}[B_2\tilde{C}_i]_T + \tilde{k}_{dps}[B_2\tilde{C}_i]_T - (\tilde{k}_{ps} + \tilde{k}_{dps})[B_2\tilde{C}_i]_T$$

$$+k_i^{Bf}\left(B_T - 2\sum_i [B_2\tilde{C}_i]_T\right)^2 \tilde{C}_i - \frac{k_i^{Bb}\tilde{K}_i\,[B_2\tilde{C}_i]_T}{\tilde{K}_i + A^m}, \quad 26$$

where the concentration of free KaiA is determined by the condition:

$$A + \sum_{i=0}^{6}\frac{AC_{i,T}}{K_i + A} + m\sum_{i=0}^{6}\frac{A^m B_2\tilde{C}_{i,T}}{\tilde{K}_i + A^m} - A_T = 0. \quad 27$$

Here $A$ and $B$ are the concentrations of free KaiA and KaiB, respectively. $C_i$ and $\tilde{C}_i$ are the free concentrations of active and inactive KaiC hexamers with $i$ monomers phosphorylated. Concentrations marked with a subscript "$T$" are the total concentrations of the respective protein, including both free protein and protein bound into larger complexes; $[B_2\tilde{C}_i]_T$ is the total concentration of inactive KaiC complexed with KaiB, including both bare BC complex and BC complex bound to KaiA.

The effective phosphorylation rates are:

$$\sigma_i^{ps} = \frac{k_{ps}K_i + k_{pf}A}{K_i + A}, \quad \sigma_i^{dps} = \frac{K_i k_{dps}}{K_i + A}. \quad 28$$

The flipping rates are $\sigma_i^{Ff} = f_i K_i / \left(K_i + A\right)$, $\sigma_i^{Fb} = b_i\,.\,K_i$ and $\tilde{K}_i$ are the dissociation constants for KaiA binding to the active and inactive allosteric state, respectively. $\tilde{k}_{ps}$ and $\tilde{k}_{dps}$ are the phosphorylation and dephosphorylation rates. The transition rates between the active and inactive allosteric states are $f_i$ and $b_i$. $K_i = k_i^{Ab} / k_i^{Af}$ and $\tilde{K}_i = \tilde{k}_i^{Ab} / \tilde{k}_i^{Af}$ are the dissociation constants for KaiA binding to the active and inactive segments of the cycle. $K_i^{Bf}$ and $K_i^{Bb}$ are the forward and backward rates for KaiB binding to inactive KaiC.

## Derivations for Monomer Model Analytics

Because of the simplicity of the monomer model it is possible to make some analytic progress by taking a limit where the phosphorylation and dephosphorylation of the $T$ residue is fast, consistent with the assumptions of the monomer model as published. This will allow us to reduce the dynamics of interest to the phase plane and use standard phase plane techniques such as nullcline analysis to estimate the dependence of the period on the sequestration stoichiometry $m$, as well as to gain a qualitative understanding of dynamics that can also be seen in the full monomer model.

## Reducing the Monomer Model

Due to the simplicity of the original monomer model it is possible to derive analytic results which can describe the relationship between $m$ and the period. The only approximation is that the separation of time scales is large. In particular, by taking phosphorylation on the threonine residue to be much faster than that on the serine residue it is possible to reduce this to a 2 degree of freedom system, for which there exist powerful tools for the analysis of nonlinear oscillations, especially of the relaxation type. This is valid because the rate constants for $T$ phosphorylation in the model as published are generally higher than those for $S$. This is implemented formally by introducing a factor of $\epsilon^{-1}$ in front of the "fast" rates (phosphorylating and dephosphorylating the threonine residue):

$$\dot{U} = \frac{1}{\epsilon}\left(k_{tu}T - k_{ut}U\right) + k_{su}S - k_{us}U, \tag{29}$$

$$\dot{T} = \frac{-1}{\epsilon}\left(k_{tu}T - k_{ut}U\right) + k_{dt}D - k_{td}T, \tag{30}$$

$$\dot{D} = \frac{1}{\epsilon}\left(k_{sd}S - k_{ds}D\right) - k_{dt}D + k_{td}T, \tag{31}$$

$$\dot{S} = \frac{-1}{\epsilon}\left(k_{sd}S - k_{ds}D\right) - k_{su}S + k_{us}U. \tag{32}$$

We then make the change of variables:

$$W = T + U, \tag{33}$$

$$X = S - D, \tag{34}$$

$$Y = S + D, \tag{35}$$

$$Z = T - U. \tag{36}$$

Solving for the new variables gives:

$$S = \frac{X+Y}{2}, \tag{37}$$

$$D = \frac{Y-X}{2}, \tag{38}$$

$$T = \frac{W+Z}{2}, \tag{39}$$

$$U = \frac{W - Z}{2}. \qquad (40)$$

Rewriting the ODE in terms of the new variables gives:

$$\begin{aligned}
\dot{W} &= \frac{Y}{2}\left(k_{su} + k_{dt}\right) + \frac{X}{2}\left(k_{su} - k_{dt}\right) - \frac{W}{2}\left(k_{us} + k_{td}\right) - \frac{Z}{2}\left(k_{td} - k_{us}\right), \\
\dot{X} &= \frac{1}{\epsilon}\left(Y\left(k_{ds} - k_{sd}\right) - X\left(k_{ds} + k_{sd}\right)\right) + \frac{W}{2}\left(k_{us} - k_{td}\right) - \frac{Z}{2}\left(k_{us} + k_{td}\right) + \frac{Y}{2}\left(k_{dt} - k_{su}\right) + \frac{Y}{2}\left(k_{dt} - k_{su}\right), \\
\dot{Y} &= \frac{W}{2}\left(k_{td} + k_{us}\right) + \frac{Z}{2}\left(k_{td} - k_{us}\right) - \frac{Y}{2}\left(k_{dt} + k_{su}\right) - \frac{X}{2}\left(k_{su} - k_{dt}\right), \\
\dot{Z} &= \frac{1}{\epsilon}\left(W\left(k_{ut} - k_{tu}\right) - Z\left(k_{ut} + k_{tu}\right)\right) + \frac{Y}{2}\left(k_{dt} - k_{su}\right) - \frac{X}{2}\left(k_{dt} + k_{su}\right) + \frac{W}{2}\left(k_{us} - k_{td}\right) - \frac{Z}{2}\left(k_{us} + k_{td}\right).
\end{aligned} \qquad (41)$$

In the $\epsilon \to 0$ limit the $\dot{Z}$ and $\dot{X}$ expressions result in self-consistent equations:

$$\begin{aligned}
0 &= -k_{tu}^{0}\left(W + Z\right) + f\left(X, Y\right)\left(k_{ut}^{A}\left(W - Z\right) - k_{tu}^{A}\left(W + Z\right)\right), \\
0 &= k_{ds}^{0}\left(Y - X\right) + f\left(X, Y\right)\left(k_{ds}^{A}\left(Y - X\right) - k_{sd}^{A}\left(X + Y\right)\right),
\end{aligned} \qquad (42)$$

where each $k_{ab}$ is of the form $k_{ab}^{0} + f\left(X, Y\right)k_{ab}^{A}$ where $f\left(X, Y\right) = \frac{\max\left\{0, A_T - m\left(X + Y\right)\right\}}{k_{1/2} + \max\left\{0, A_T - m\left(X + Y\right)\right\}}$.

Note that since $X$ is bistable it cannot be assumed to be at its equilibrium. When there is no free KaiA, $f\left(X, Y\right) = 0$ and the expression for $X$ and $Y$ simply gives $X = Y$. Then when there is free KaiA we can take $f\left(X, Y\right) = \frac{A_T - m\left(X + Y\right)}{k_{1/2} + A_T - m\left(X + Y\right)}$. The equation for $X$ and $Y$ gives:

$$\begin{aligned}
0 = \; & X^2 m\left(k_{ds}^{0} + k_{sd}^{A} + k_{ds}^{A}\right) \\
& + X\left(-k_{ds}^{0}\left(k_{\frac{1}{2}} + A_T - mY\right) - mk_{ds}^{0}Y - \left(k_{ds}^{A} + k_{sd}^{A}\right)\left(A_T - mY\right) + mY\left(k_{ds}^{A} - k_{sd}^{A}\right)\right) \\
& + k_{ds}^{0}Y(k_{1/2} + A_T - mY) + \left(A_T - mY\right)\left(k_{ds}^{A} - k_{sd}^{A}\right)Y,
\end{aligned} \qquad (43)$$

### Finding the Turning Points in the Reduced Monomer Model

One can see from this expression that there will be terms sublinear in $Y$ in the expression of $X(Y)$. This also provides the endpoints of each branch. One endpoint is found by finding the intersection of this expression with the line $X = Y$, the expression for the nullcline when there is no free KaiA. This reduces to:

$$0 = 2Y^2 m + Y\left(-A_T\right), \quad 44$$

or

$$Y = \frac{A_T}{2m}. \quad 45$$

To find the other turning point we must remember that the parabola described above must have two distinct branches as a function of $X$ as a function of $Y$. We can then find the point at which the two branches meet. This occurs at the point where the discriminant of the quadratic equation for $X$ is equal to $0$. This requirement is expressed by the relation:

$$\begin{aligned} 0 = & -\left(-k_{\frac{1}{2}} k_{ds}^{0} - A_T\left(k_{ds}^{0} + k_{ds}^{A} + k_{sd}^{A}\right) + 2k_{ds}^{A} Y m\right)^2 \\ & +4m\left(k_{ds}^{0} + k_{ds}^{A} + k_{sd}^{A}\right)\left(mY^2\left(-k_{ds}^{0} + k_{sd}^{A} - k_{ds}^{A}\right) + k_{ds}^{0} Y\left(A_T + k_{\frac{1}{2}}\right) - YA_T\left(k_{sd}^{A} - k_{ds}^{A}\right)\right). \end{aligned} \quad 46$$

This expression is now quadratic in $Y$, giving two solutions:

$$\begin{aligned} 2\left(k_{ds}^{0} + k_{ds}^{A}\right)^2 mY = \; & A_T\left(k_{ds}^{0} + k_{ds}^{A}\right)\left(k_{ds}^{0} + k_{ds}^{A} + k_{sd}^{A}\right) + k_{\frac{1}{2}} k_{ds}^{0}\left(k_{ds}^{0} + k_{ds}^{A} + 2k_{sd}^{A}\right) \\ & \pm 2\left[k_{\frac{1}{2}} k_{ds}^{0} k_{ds}^{A}\left(k_{\frac{1}{2}} k_{ds}^{0} + A_T\left(k_{ds}^{0} + k_{ds}^{A}\right)\right)\left(k_{ds}^{0} + k_{ds}^{A} + k_{sd}^{A}\right)\right]^{1/2}. \end{aligned} \quad 47$$

Obviously only one of these can be the actual value at which the points of the parabola meet so by plugging in the published parameter values it is clear that the negative of the radical must be taken since if the positive term is taken the value achieved is several orders of magnitude too high (approximately $1450/m$ as opposed to approximately $1/m$).

### Estimating the Period Dependence of the Monomer Model

By reducing the system to 2 degrees of freedom, $X$ and $Y$, we obtain a model that is amenable to phase plane analysis. In Figure 9e we plot the nullclines of this system, curves along which each degree of freedom is constant. These nullclines have a motif, one nullcline has an “S” shape and the other has a linear section that crosses the middle branch of the “S”-shaped nullcline, that indicates a type of positive feedback dynamics known as a relaxation oscillator [26]. This describes a system which tracks slowly along the outer branches of “S”-shaped nullcline until the vertical cusp is reached. It then quickly switches to the other branch and moves slowly in the opposite direction until it reaches the other cusp and switches back to the first branch, restarting the cycle. Relaxation oscillators are sometimes understood in terms of simple electric circuits which involve a capacitor slowly charging up to a certain voltage, suddenly discharging, and the slowly building a charge back up again. It appears that a similar mechanism is at work in the KaiA sequestration dynamics of this model.

The linear part of the of the “S”-shaped nullcline corresponds to the situation in which all of the KaiA is sequestered and the KaiC monomers are dephosphorylating, in which case the original model becomes completely linear. This can be seen directly from the nullcline plot as both $X$ and $Y$ are decreasing indicating that the sum $S+T$ is decreasing, indicating dephosphorylation, but the relative amount of $S$, the sequestering protein, is also decreasing. By reducing the model to a one dimensional model along this nullcline it is possible to derive an analytic expression for the time spent on this branch as a function of $m$. The amount of time the system spends with all KaiA sequestered, which we call $T_{seq}$, is of the form:

$$T_{seq} = \int_{\frac{z_1}{m}}^{\frac{z_2}{m}} \frac{1}{\dot{Y}_{seq}(X(Y),Y)} dY = \int_{\frac{z_1}{m}}^{\frac{z_2}{m}} \frac{1}{-cY} dY = -\frac{\ln(Y)}{c}\Big|_{\frac{z_1}{m}}^{\frac{z_2}{m}} = -\frac{\ln(z_2)-\ln(z_1)}{c}, \qquad 48$$

Where the values for $z_i$ and $c$ are given by the expressions for $\dot{Y}$ and the turning points above. This result is notable in that it suggests that even in the full model the time spent fully sequestered does not vary strongly with $m$. This is supported by Figure 7b. The expression for the time derivative of $Y$ is linear since for zero free KaiA even the full three degree of freedom system becomes linear. The turning points can be shown to depend only on the reaction rate constants and the total KaiA concentration.

On the other branch the KaiC monomers are phosphorylating, first on the $T$ residue and then more slowly on the $S$ residue. This is can be seen on the nullcline plot by noting that on the nonlinear branch $Y$ increases but $X$ remains roughly constant, actually decreasing slightly, indicating an increase in the doubly phosphorylated KaiC as well as a smaller increase in $S$ phosphorylated KaiC. Then once a certain amount of KaiA has been sequestered the doubly phosphorylated KaiC begins to dephosphorylate, increasing the amount of $S$, causing more KaiA to be sequestered in a positive feedback process. The KaiC monomers then begin to dephosphorylate and the cycle begins again. The time spent when the amount of free KaiA is nonzero has the form:

$$T_{unseq} = \int_{\frac{z_1}{m}}^{\frac{z_2}{m}} \frac{1}{\dot{Y}_{unseq}(X(Y),Y)} dY. \qquad 49$$

$\dot{Y}_{unseq}$ is much more complex so the integral is not directly tractable. Despite this it is possible to gain some insight from this expression. In this case $X(Y)$ is the solution to a self-consistent equation that leads to a quadratic equation for $X$, of the form:

$$X(Y) = \frac{d_1 + d_2 mY \pm \left(d_3 + d_4 mY + d_5 m^2 Y^2\right)^{1/2}}{m}. \qquad 50$$

The sublinear term in $X(Y)$ causes $\dot{Y}_{unseq}$ to vary sublinearly with $Y$ and integrating such a term between endpoints $z_1$ and $z_2$ would cause the value of the integral to vary inversely with $m$.

## Materials and Methods

ODE integration was performed in MATLAB using the ode15s integrator. The nullclines for the monomer model were calculated in Mathematica 11.1 using the ContourPlot function. The 2-state allosteric model was evolved using a kinetic Monte Carlo algorithm described in [32].

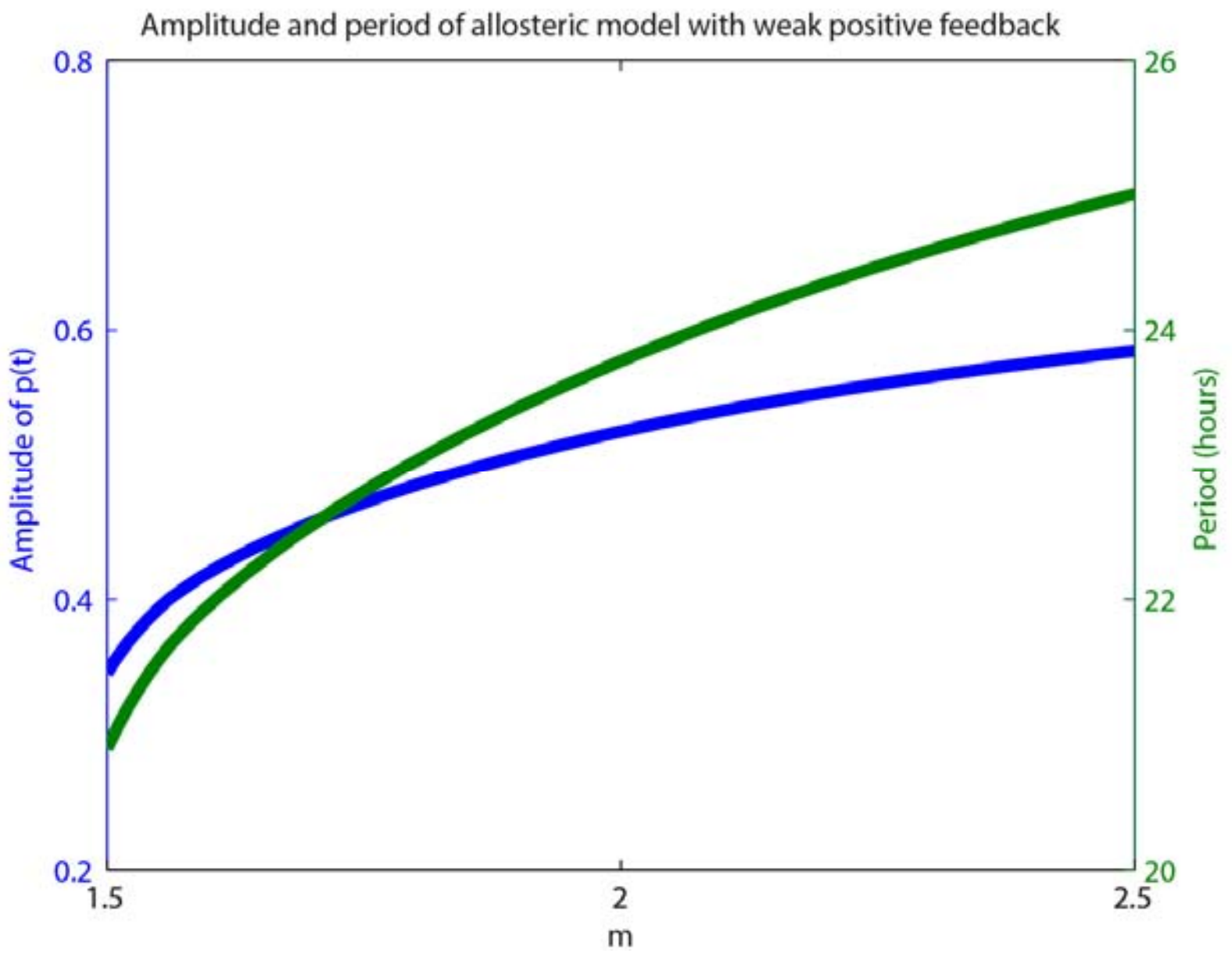

Figure 13

The qualitative trends in the amplitude and period dependence of the allosteric model are robust to the addition of a small amount of positive feedback into the model. Parameters are identical to the standard allosteric model but with KaiA binding to the C-terminal domain and promoting phosphorylation on inactive KaiC hexamers at 1/100 the affinity of an active hexamer

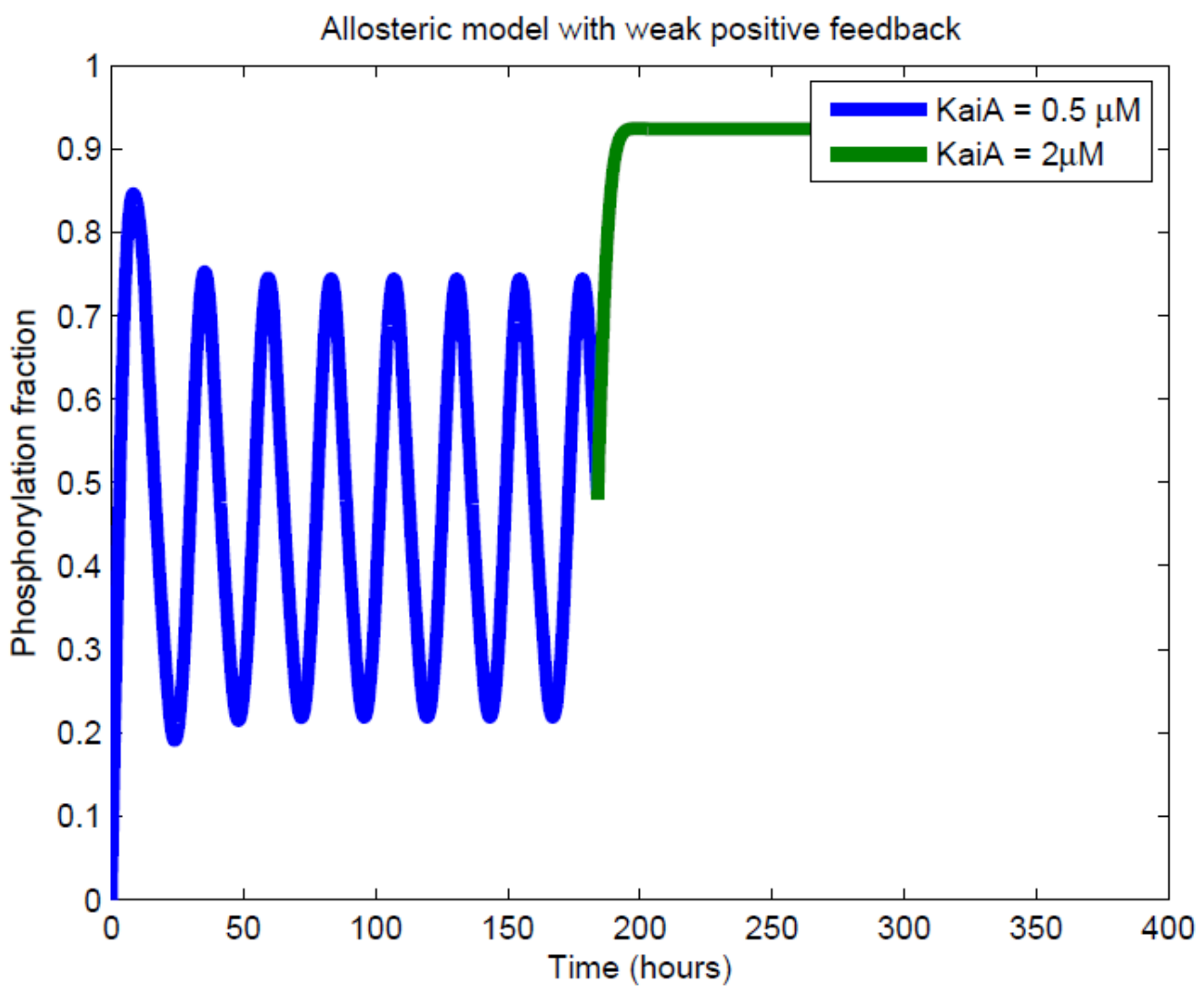

Figure 14

The addition of a small amount of positive feedback to the allosteric model allows it to reproduce the experimental result that adding a large amount of KaiA during dephosphorylation will cause an increase in phosphorylation, while maintaining dynamics that are generally dominated by negative feedback and delay effects. The addition of KaiA is indicated by the trace changing from blue to green.

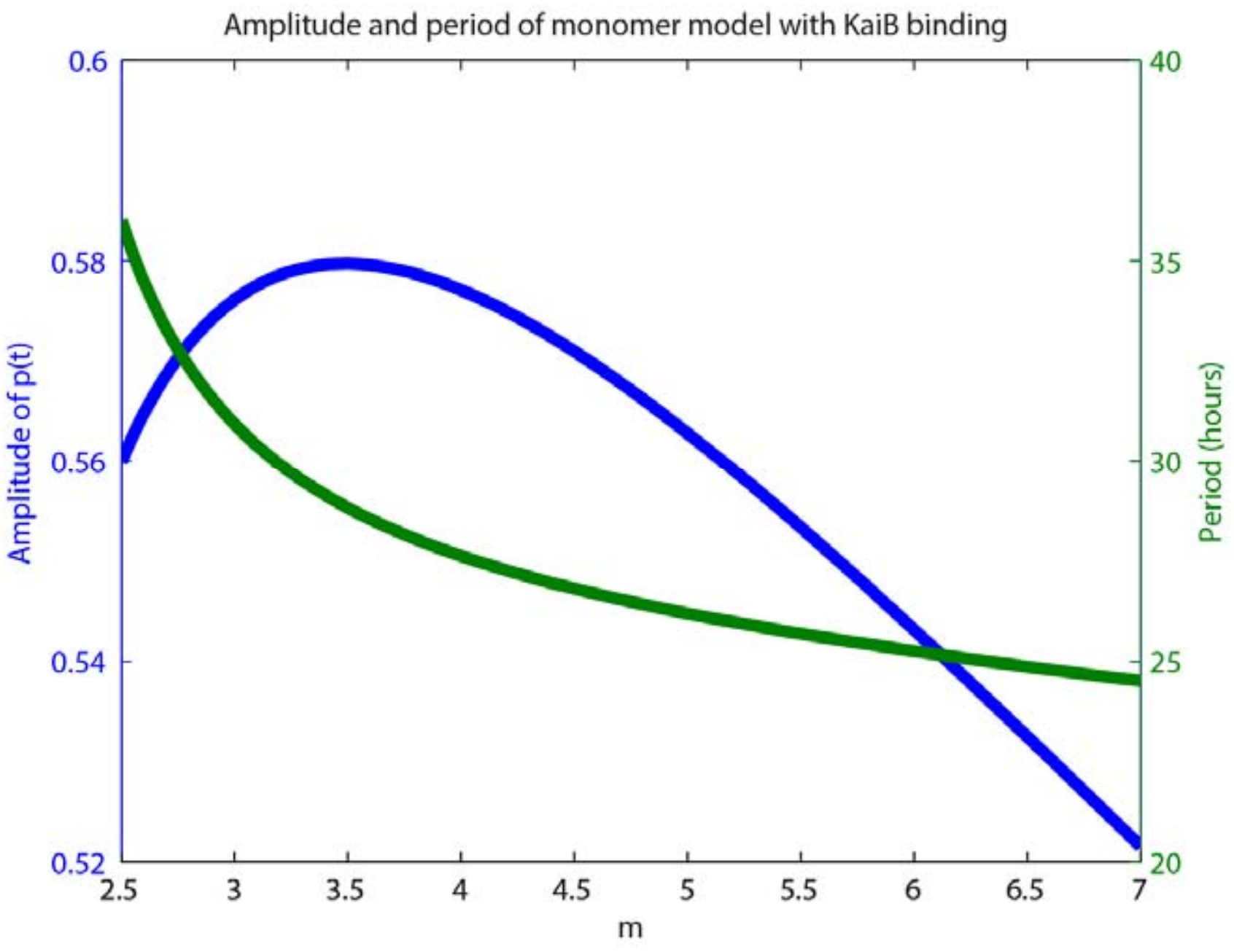

Figure 15

The amplitude and period of the monomer model with explicit KaiB binding as described show the same general trends and features as in the original monomer model.

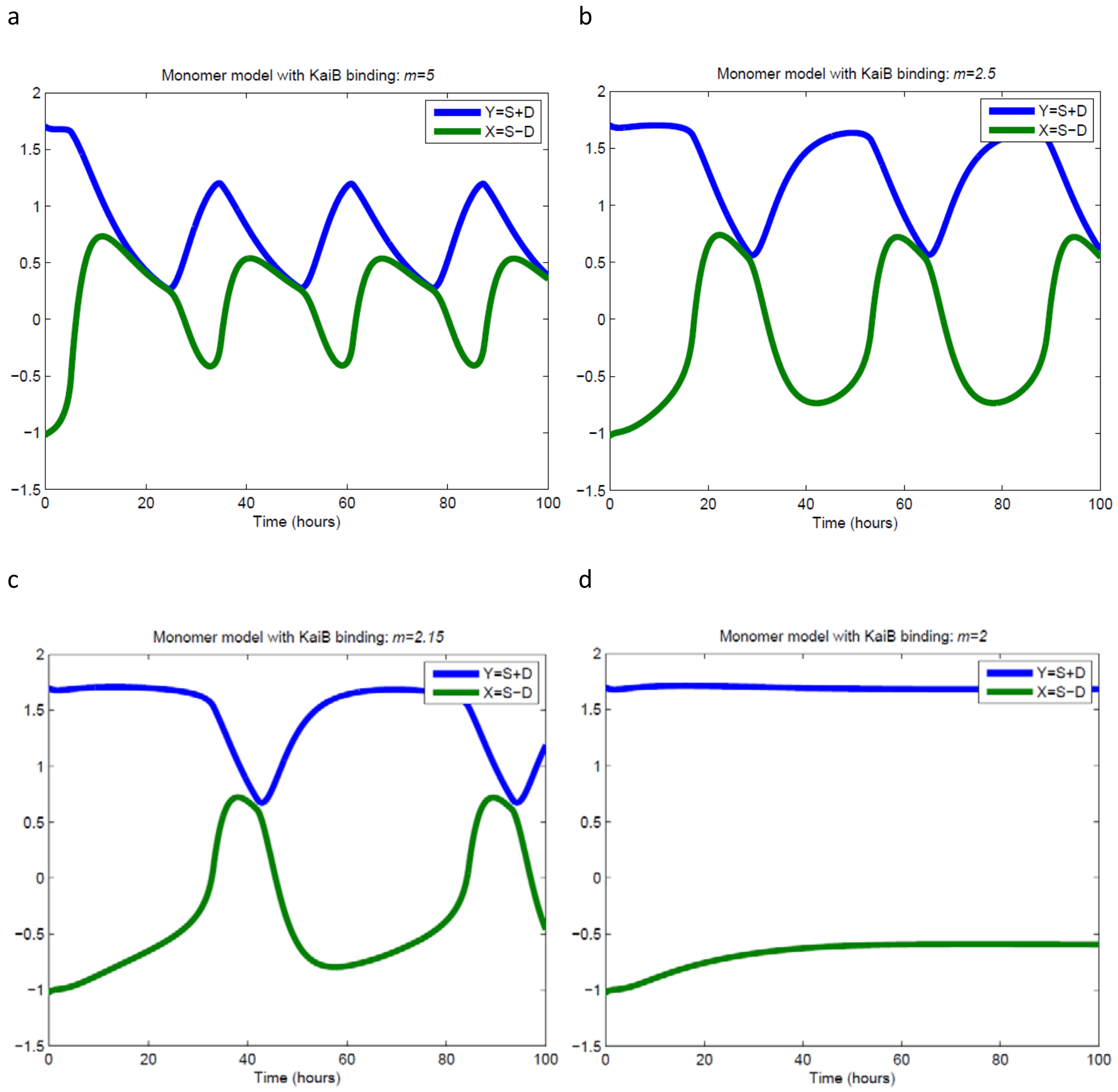

Figure 16

An examination of a version of the monomer model extended to include explicit KaiB binding. Comparing to Fig. 10, it is possible to see the same general qualitative features that were predicted by the reduced 2 degree of freedom model, as described in the main text.

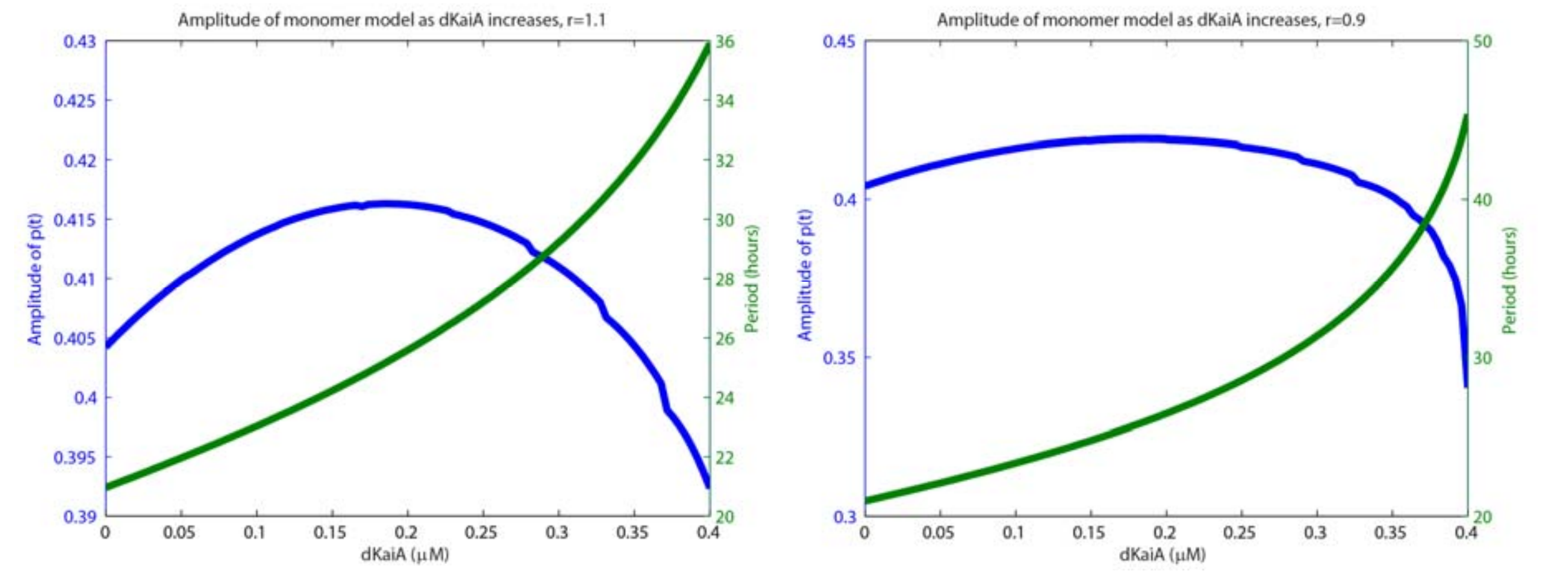

Figure 17

Amplitude and period for the original monomer model with decoy KaiA that binds with a different affinity than KaiA. (a) Decoy KaiA is sequestered with affinity $r = 1.1$ times the affinity of KaiA. (b) Decoy KaiA is sequestered with affinity $r = 0.9$ times the affinity of KaiA.